**Accounting for shelf-width in selecting altimetry observations for coastal sea level variability improves its agreement with tide gauges**

**Vandana Sukumaran[1] and Bramha Dutt Vishwakarma[1,2]**

[1]Interdisciplinary Centre for Water Research, Indian Institute of Science, Bangalore – 560012,

[2]Centre for Earth Sciences, Indian Institute of Science, Bangalore – 560012

**Abstract**

A novel dynamically varying search radius algorithm is developed that takes advantage of bathymetry information to choose satellite observations that represent coastal sea level variability better. The algorithm is successfully tested at 155 tide gauge stations around the globe and demonstrates broader applicability across different coastal regimes compared to existing validation methods. This is supported by consistently higher median correlation and lower median Normalized Root Mean Square Error (NRMSE). Furthermore, the new algorithm improves the efficacy of the low-resolution product, X-TRACK SLA L2P v2022 (XTRACK), and makes it comparable to the high-resolution (20Hz) coastal products: Along-track sea level anomalies and trends v2.3. Using the algorithm at over 267 stations, XTRACK data shows improved agreement with tide gauges for both linear and non-linear trends. In some regions, such as tidally-dominated estuaries and the Eastern Australian coast, lower correlation and higher RMSE for residual signals are reported, which are discussed.

## Main

Sea level (SL) rise will be one of the most disastrous consequences of anthropogenic climate change. About 40% of the world's population (~2.5 billion people) lives within 100km of the coast and would be severely affected[1–3]. Hence, monitoring and understanding of coastal SL change is crucial. Coastal SL variability differs from the open ocean due to the influence of continental slope, shallow waters, and the presence of coastlines[4–7]. Tide gauges (TG), the oldest and the direct source of coastal SL data, are sparse, unevenly distributed, and suffer from data gaps[8,9]. Satellite altimetry has monitored global SL since 1992, but its reliability near the coast remains limited[10]. Closer to the coast, altimetry signals are affected by bright land returns[11,12], off-nadir returns from calm water[11–13], steep waves[14,15], and short-scale variability of surface characteristics[11,12,16]. Off-nadir returns from calm-water contaminate the waveform[15–17], while bottom friction in shallow waters produce steep waves which cause a negative bias in significant wave height (SWH) when compared with buoys[18]. Small-scale waves reduce nadir backscatter[15], and short-scale water vapor variations affect the wet-tropospheric correction[19]. Hence, observations closer than 50km to the coast were neglected until 2010[20,21]. Several efforts have been made to retrieve coastal SL signals from different altimetry missions: PISTACH for Jason (2007–2011), COASTALT for Envisat (2008–2012), X-TRACK, PEACHI for AltiKa, ALES, SARvatore for CryoSat-2 and Sentinel-3, SARINvatore for CryoSat-2, SAMOSA+ and SAMOSA++ for SAR altimeters[22,23]. Recently, fully focused SAR (FF-SAR) processing has been found to yield sub-meter along-track resolution and more accurate geophysical estimates (like sea surface height (SSH) and SWH) since it is less affected by strongly reflective targets near the coast[24].

X-TRACK is a dedicated post-processing system for coastal zones that enhances both the quantity and quality of SSH observations[25]. ALES (Adaptive Leading Edge Subwaveform) retracker derives SSH from coastal and open ocean observations by improving the range term[26]. The ALES algorithm has been extensively validated in numerous studies[26–28]. For example, Marti et al. used Jason-1,2 altimetry data (ALES retracked and X-TRACK post-processed) along the Western African coast, using the along-track pass closest to the shoreline, and found coastal sea level trends (CSLT) stable between 15–5km offshore but varying nearshore[29]. Similarly, Gouzenes et al. reported good agreement of CSLT beyond 4km along the coast of Senetosa[30]. Abele et al. evaluated the performance of ALES and SAMOSA++ using SL variance in the coastal zone of the Northwest Atlantic and found that SAMOSA++ outperformed ALES in the coastal strip (<10km from the shore)[31].

Altimetry and TG observations are not co-located and thus, several methods have been used to aggregate altimetry data prior to validation[13,32–34]. Vinogradov and Ponte compared annual amplitudes and phases at 345 continental and island stations with altimetry, from TOPEX/Poseidon, using a search radius (SR) of 134km centered on TG[34]. They found that the correlations of annual amplitudes decrease with increasing depth, and an average r=0.85 in nearby shallow regions. In a separate study, they compared the low-frequency coastal sea level anomaly (SLA) variability using an SR of 180km. They reported the best agreement for Pacific and Indian

Ocean islands and Western Australia, but large discrepancies (>4cm RMSE) along southwest America, Japan, and Eastern Australia[35]. Birol et al. compared a new high-resolution coastal altimetry product called XTRACK/ALES (20Hz) with XTRACK (1Hz and 20 Hz) across several regions (except Northwest and Northeast America, South America, and Eastern Africa). The product uses Jason 1,2,3 measurements retracked with ALES and post-processed with X-TRACK software. They compared the nearest altimetry point (having 80% time-series (TS) completeness) with TG, and found that XTRACK/ALES, after filtering, shows slightly higher correlation ($r \geq 0.7$) and better Root Mean Square Error (RMSE) than XTRACK 1Hz at all stations except Sete [36]. Cazenave et al. validated monthly SLA from point-level XTRACK/ALES observations and found an average correlation of 0.5, which improved to 0.78 when altimetry was averaged along and across tracks near each TG, due to reduced sampling errors and influence of small-scale variability[37]. Linear trends also agreed statistically at 64% of stations[37].

All these studies are either regional or concerned with certain temporal scales (interannual, annual, trend, residual). A comprehensive global assessment at various temporal scales is yet to be carried out. The coastal SL variability is more sensitive to local ocean dynamics, which is influenced by the continental shelf-width[4,38,39]. Thus, selecting altimetry points along the coast and within the shelf would represent the coastal SL better. Hence, we develop a novel Dynamic Search Radius (DSR) algorithm, which selects altimetry observations along the coast but within the shelf-break, and apply it on X-TRACK SLA L2P v2022, 1Hz (hereinafter XTRACK). An exhaustive validation and assessment of this algorithm is performed against TG, and a high-resolution (20Hz) coastal altimetry product (Along-track sea level anomalies and trends v2.3 product (2023), hereinafter XTRACK/ALES). The new DSR is evaluated against six existing algorithms applied to XTRACK and five applied to XTRACK/ALES, and is validated across 155 TG stations globally. In addition, the capability of DSR to capture SLA signals is evaluated at various temporal scales: non-linear trend, seasonal, and residual at 287 stations. Finally, we investigate CSLT at 267 stations and analyze regional patterns.

## The Dynamic Search Radius algorithm

The shelf-width and its gradient play a dominant role in coastal SL dynamics, thus including this information appears to be vital in the selection of altimetry observations, but is not considered in the existing algorithms. Hence, a new selection algorithm is proposed that considers altimetry observations falling inside the shelf-break. Shelf-break is defined as the location where the continental shelf meets the continental slope. Although found to exhibit a wide variability, it is commonly defined at the 200m isobath[40–55] or at the 34.5 isohaline[56,57]. Several studies across various disciplines approximate the shelf-break at 200m isobath[41,42,44,46–55]. For example, Graham et al. assessed the strength of downwelling circulation around the European northwest shelf across the shelf-break defined at 200m isobath[41], while Franco et al. showed long-term trends of sea surface temperature, chlorophyll-a, and absolute dynamic topography in the Patagonian shelf-break front, also along the 200m isobath from GEBCO[48]. Hence, the shelf-break in this study is defined at the 200m isobath.

In the first step of the algorithm, points along the coastline are selected, extending 15km (parameter #1, called as coastline extent) in either direction from the TG location. Next, the maximum width of the shelf-break within coastline extent is calculated and designated as the first SR (SR1) for that TG. The mean of altimetry observations inside this SR1 represents the coastal SLA. While aggregating, altimetry points with less than 80% completeness in their TS are excluded[36,58,59]. Further, we set a lower limit of 15 altimetry points to be present within SR1; defined as the threshold (parameter #2). If this condition is not satisfied, the SR is updated to 250km, defined as SR2 (parameter #3), and altimetry points only inside the shelf-width (maximum shelf-width (value of SR1)) are selected. Then the mean SLA from the selected observations is calculated. If the threshold is still not satisfied, we believe that not many observations are available to find altimetry SLA, and we discard that station from further analysis. Since the width and depth of shelf is found to be dramatically varying along the coastline[60], we use the largest width of the shelf-break in the 30km coastline (since parameter #1 =15km), which allows us to capture most of the observations along the coast but within the shelf over that region. The values of the three parameters are chosen through a sensitivity analysis (SA) spanning 396 parameter combinations across 318 TG stations (see Methods, Supplementary Fig.S1-S3). Since the SR is dynamic and is decided for each TG based on the nearby shelf-break, it is called the Dynamic Search Radius algorithm. An illustrative explanation of the algorithm is shown in Fig.1, and the flowchart is in Supplementary Fig.4.

This algorithm is not applied on TG installed on minor islands (table of minor islands is provided in Supplementary Table S1) as they are primarily surrounded by open ocean (depths>200m) and the coastal shelf is very narrow. However, to include minor islands in the global assessment of CSLT, a fixed SR of 200km is chosen, to calculate the mean of all the altimetry observations for that island station.

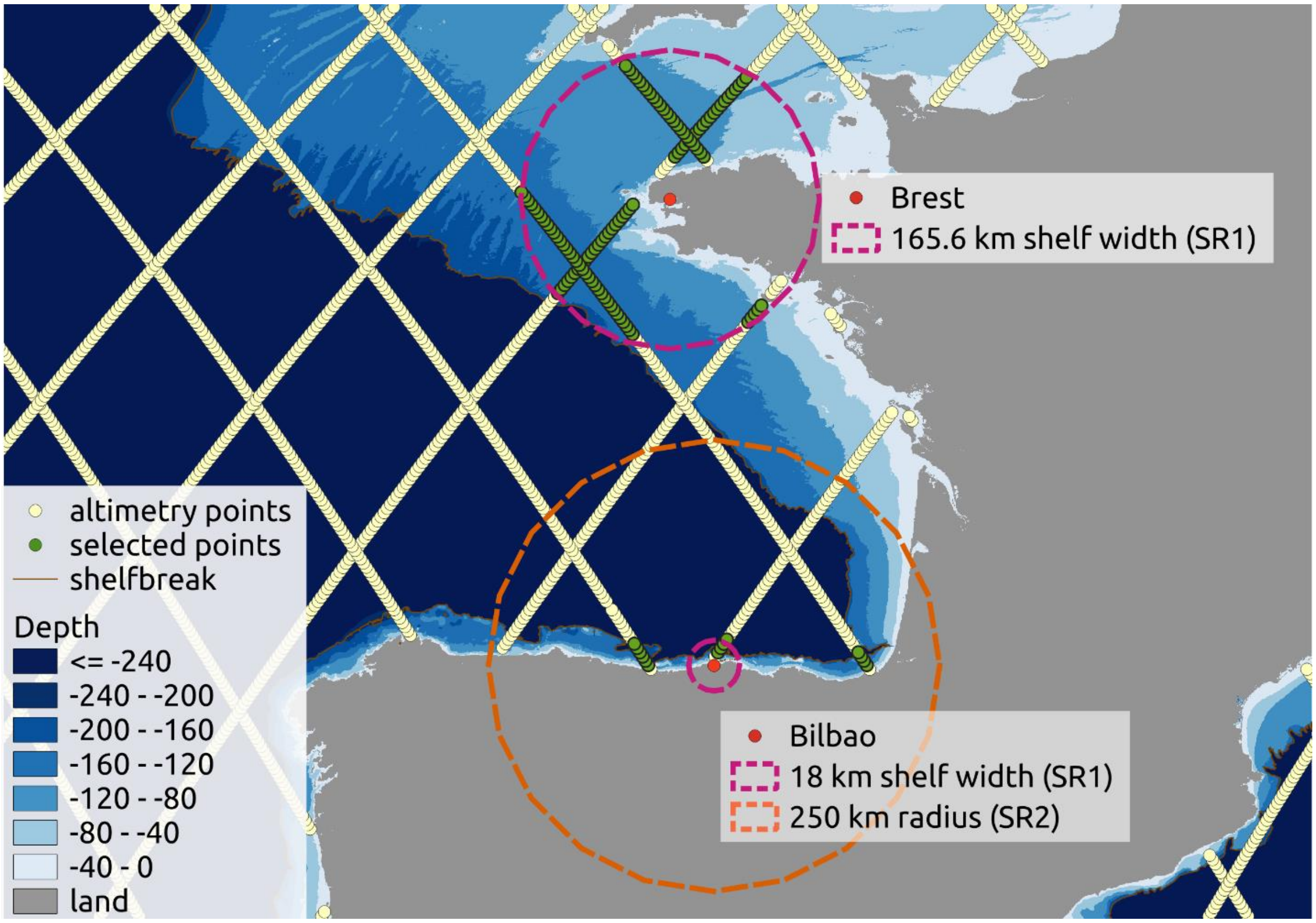

**Fig.1:** Illustrative example of DSR. At station Brest (station ID-1), SR1 is 165.6km which is the maximum shelf-width along the coastline extent of 15km from the station. Here, upon satisfying the criteria of threshold (15 observations), all observations inside SR1 are selected (green circles). At station Bilbao (station ID-1806), due to insufficient observations (less than threshold) with SR1 (18km), SR2 of 250km is used and satellite observations only within the coastal shelf is selected (green circles). Red circles represent the TG and light-yellow circles are altimetry observations. Note that some altimetry points although within the shelf are not selected, since they do not meet the 80% completeness criterion.

**A comprehensive validation and assessment of coastal altimetry**

The novel DSR, along with other popular approaches, such as fixed SR, nearest pass, and nearest observation, is compared against TG observations. To implement the fixed SR algorithm, the mean of altimetry observations within a chosen SR centered at the TG is taken as the corresponding coastal SL. We used 150km (XTRACK: X_150), 200km (XTRACK: X_200, XTRACK/ALES: ales_200), 250km (XTRACK: X_250), and 300km (XTRACK: X_300) as fixed SR. In the nearest observation algorithm, the nearest satellite point having at least 80% completeness [36,58,59] in the SLA is chosen (XTRACK: X_npoint, XTRACK/ALES: ales_npoint). In the nearest pass (nearest virtual station) and nearest two-pass algorithm, the mean SLA of observations from the nearest pass (XTRACK/ALES: ales_np) or nearest two passes (XTRACK/ALES: ales_2np) is chosen. In the 250km nearest pass method, altimetry points from the nearest pass that are within a 250km radius from the TG are selected (XTRACK: X_np250). The XTRACK/ALES product provides high-resolution 20Hz SLA from 20km offshore to the coast[61]. On narrow land stretches and peninsulas, a 300km SR may pick observations from the opposite side of the land. This has been curtailed to the extent possible by only taking mean of observations from the nearest site (station-number) of a pass (pass-number) (XTRACK/ALES: ales_300).

For intercomparison, 155 TG stations are selected based on the following criteria: TG data available from 2002 with complete coverage for 2002–2020, ≤20% total data gaps (excluding any ≥4 months continuous gaps), and availability of satellite observations within 100km from both XTRACK and XTRACK/ALES. The performance of altimetry against TG is assessed using r, RMSE, and Normalized Root Mean Square Error (NRMSE). An altimetry observation is considered to exhibit good agreement with TG data if the r exceeds 0.7, or the RMSE is below 20mm, or the NRMSE is below 1. The XTRACK/ALES product provides deseasonalized SLA. Hence, to facilitate the intercomparison of XTRACK with TG and XTRACK/ALES, annual and semi-annual signals are removed. The detrended deseasonalized (DeST) SLA are then compared, with the linear trend being removed using least-squares regression. The stations are grouped based on their coastal region: Gulf of Mexico (GoM), Indian Ocean (IO), Mediterranean Sea (MedSea), Northeast Atlantic and North Sea (NEAt-NS), North Pacific Ocean (NPO), Northwest Atlantic (NWAt), South China Sea (SCS), South Pacific Ocean (SPO), and Southwest Atlantic (SWAt) as shown in Fig.2(c). Box plots of r and NRMSE are shown in Fig.2 and Extended Data Fig.1, respectively.

DSR consistently outperforms other algorithms across most regions, with median r(NRMSE) values of 0.81(0.77) for GoM, 0.89(0.49) for IO, 0.81(0.78) for MedSea, 0.73(0.77) for NEAt-NS, 0.85(0.61) for NPO, 0.72(0.75) for NWAt, 0.7(0.78) for SCS, 0.7(0.86) for SPO, and 0.33(0.96) for SWAt. All XTRACK algorithms perform well in IO, with X_DSR showing the highest median r of 0.89. The XTRACK/ALES-based algorithms, ales_200 and ales_300, also demonstrate comparable performance, each yielding an r=0.89. Among the algorithms, ales_npoint performs the worst across all nine regions (Fig.2(b)), while X_npoint shows the poor performance in most regions, except IO (Fig.2(a)). In NPO, all XTRACK/ALES algorithms, except ales_npoint, perform well (Fig.2(b)). The narrow shelf in this region may explain why retrieving observations closer to the coast improves performance, with ales_300 achieving a median r=0.84. Notably, despite its lower-resolution, the X_DSR algorithm delivers comparable performance, achieving the highest median correlation of 0.85 in this region. Among all the XTRACK/ALES algorithms, ales_300 performs the best. The nearest virtual station from XTRACK/ALES (ales_np) falls behind X_DSR across all regions, showing a correlation of 0.62(0.81 for X_DSR) in GoM, 0.83(0.89) in IO, 0.66(0.81) in MedSea, 0.5(0.73) in NEAt-NS, 0.8(0.85) in NPO, 0.54(0.72) in NWAt, 0.43(0.7) in SCS, 0.5(0.7) in SPO and 0.23(0.33) in SWAt.

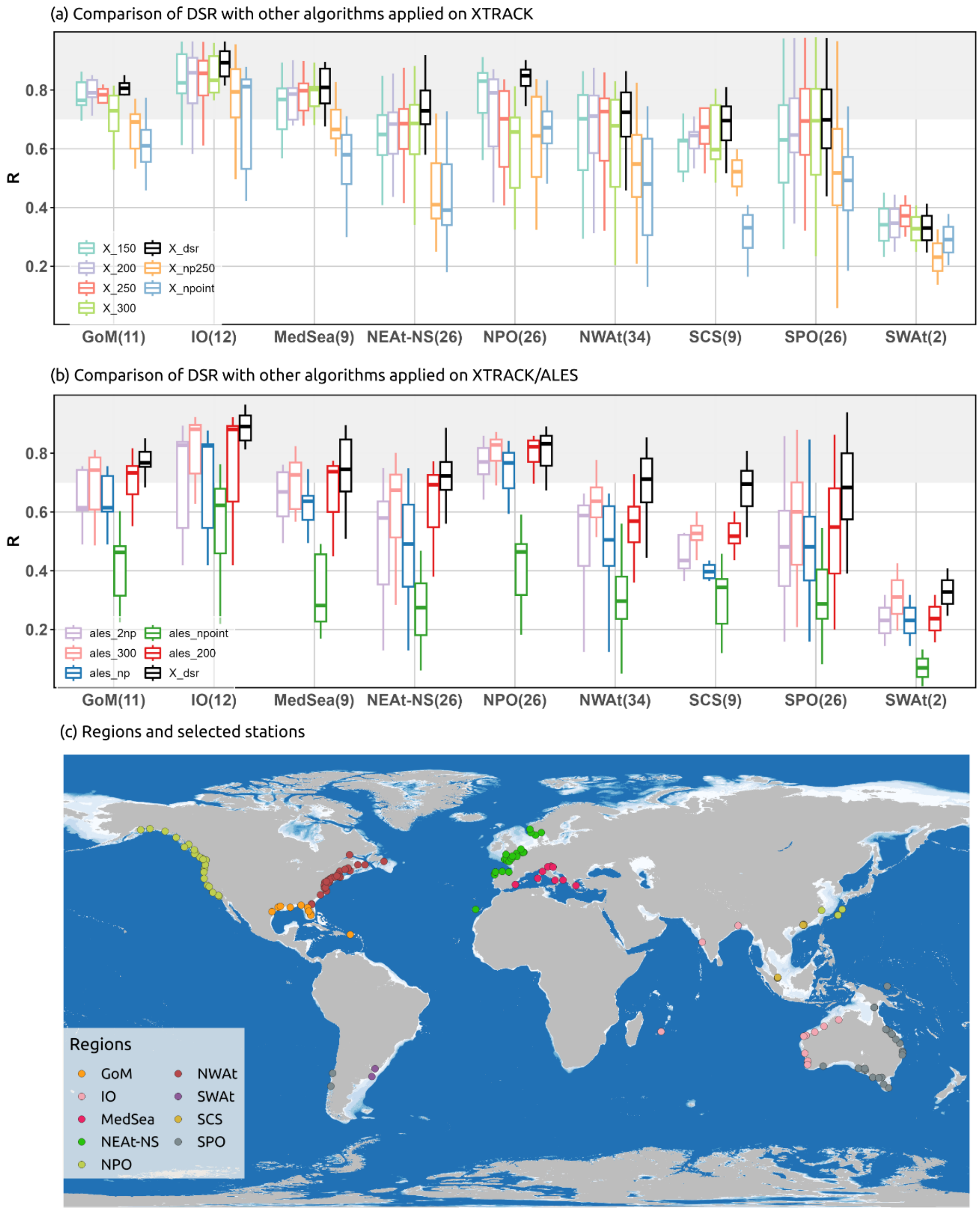

**Fig.2:** Evaluation of different algorithms for representing coastal SL variability. Fig.2(a) shows DSR (X_DSR) for XTRACK product compared with six alternate algorithms applied on XTRACK: X_150, X_200, X_250, X_300, X_np250 and X_npoint. Fig.2(b) shows DSR (X_DSR) compared with five alternate algorithms applied on XTRACK/ALES: ales_2np, ales_300, ales_np, ales_npoint, ales_200. The comparison is based on r and is conducted at 155 TG stations. Box plots of region-wise r is shown, with r ≥ 0.7 considered as good agreement, shaded in grey. Stations from different regions are shown as colored circles, with each color representing a specific region.

**Comparing XTRACK/ALES and XTRACK with TG data**

Among the 155 stations analyzed, 106(~68%) show good correlation with XTRACK (X_DSR) (Fig.3(a)), compared to 72(~46%) with XTRACK/ALES (ales_300) (Fig.3(b)) in capturing DeST SLA. Based on r, XTRACK outperforms XTRACK/ALES at 38 stations, whereas XTRACK/ALES exceeds XTRACK at only 4, and even then, the improvement is marginal (e.g., at Villagarcia, r(RMSE) is 0.74(29mm) for ales_300 and 0.66(32mm) for X_DSR).

The high-resolution XTRACK/ALES product (20Hz) performs better at narrow coasts due to its ability to retrieve observations closer to shore. However, the DSR algorithm now enables even the low-resolution (1Hz) XTRACK product to perform at par in these regions. Examples include Port Giles, Port Rupert, Sandy Hook, and most stations along the Northwest America (NWA) coast. The strict editing applied during XTRACK/ALES processing results in the exclusion of some observations[36], leaving some regions—such as the Belgian coast, without data. This explains the poor performance at stations located here, but even with the availability of observations retrieved close to the TG, several stations located along the English Channel, the St. Vincent Gulf (Australia), and the Chesapeake Bay to the Bay of Fundy (except for Long Island Sound) continue to show very poor r. For example, even though XTRACK/ALES data is available at ~30km from TG, r(RMSE) at Devonport from ales_300 is 0.49(33.7mm), whereas from X_DSR it is 0.78(21.5mm). Another example is Trieste; altimetry observations from both are available as close as 31km. The nearest XTRACK/ALES point (31.2km) shows an r of 0.52, while the nearest XTRACK point shows a slightly higher r of 0.6. All XTRACK/ALES observations have r <0.6, whereas XTRACK ranges from 0.5 to 0.78, with the highest (0.78) observed at a point 134.8km away. Altimetry observations within the Gulf of Trieste show lower r for both products, though XTRACK performs slightly better, likely due to the averaging in the 1Hz product. In contrast, observations outside the Gulf of Trieste show stronger correlations, as seen in the XTRACK product, which explains its superior performance at this location (see Supplementary Fig.S7-S8). Likewise, at the Helgeroa station, which has a narrow continental shelf (~16km), observations in shallow waters, despite being farther from the TG, capture the DeST signal more effectively than the nearest altimetry pass that falls in deeper waters (see Supplementary Fig.S9-S10).

The DeST represents the residual signal, which includes low-frequency signals such as interannual and various monthly to interannual signals. While DeST may still retain signatures of storm surges, these are likely to be significantly damped at monthly scales. Instead, the remaining variability (excluding the influence of atmospheric pressure) probably reflects processes such as SLA driven by local-wind forcing, remote forcing that induces mass fluxes, and SL variability due to coastal trapped waves (CTW)[63]. We observe that stations located in tidally-dominated estuaries, such as the Bay of Bengal (Haldia), Oujiang Estuary (Kanmen), Rio de la Plata Estuary (Mar del Plata, Montevideo), Bay of Fundy (East Port, Yarmouth), and Pearl River Estuary (Tsim Bei Tsui, Shek Pik etc.), exhibit poor correlation. In contrast, stations located in river-dominated estuaries, like the Mississippi River Delta, show good correlation. Stations located in the Gulf of Lawrence generally exhibit poor correlation. Observations at stations on the western coast of Australia demonstrate stronger correlation and lower RMSE; however, along the eastern coast, we find lower correlation and higher RMSE. Notably, altimetry fails to capture the DeST signal, even for observations taken within the shelf-break and as close as 10km from the TG. Similar findings are reported in a recent study by Benveniste et al.[62]. This discrepancy can be attributed to the Eastern Australian Current (EAC), which flows near the coast and is associated with highly variable mesoscale EAC eddies[62].

We find good agreement with TG from X_DSR at Long Island Sound situated between Chesapeake Bay and Bay of Fundy, Western and Southern Australia, Norwegian coast, Strait of Juan de Fuca, Gulf of Mexico, NWA, and North Sea beyond the Strait of Dover.

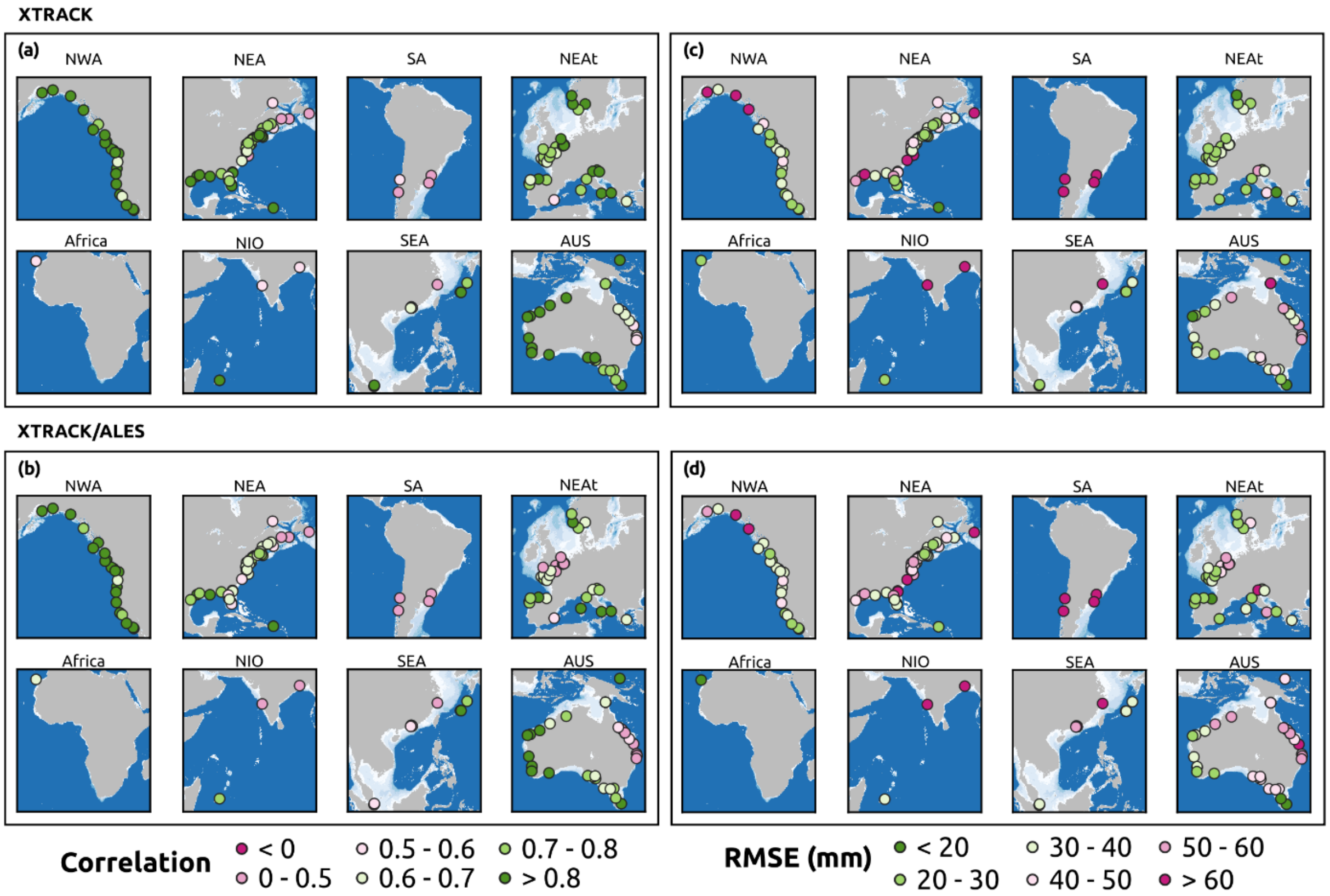

**Fig.3:** Validation of XTRACK (X_DSR) with XTRACK/ALES (ales_300): Correlation and RMSE with TG. Circles represent TG stations, with circle color indicating the correlation and RMSE. Shades of pink indicate poor correlation and large RMSE, while shades of green represent good correlation and small RMSE. (a) and (c) show correlation and RMSE of XTRACK with TG, while (b) and (d) show correlation and RMSE of XTRACK/ALES with TG.

### XTRACK's performance in capturing SL variability at different temporal scales

The SLA from XTRACK and TG are first decomposed into Non-Linear Trend (NLT), annual (seasonal), and residual signals, filtered using the DSR algorithm, and then compared. NLT includes low-frequency signals such as interannual variability and linear trend; the residual signal captures monthly to interannual signals; and the annual signal represents the seasonal variability in SL. The analysis is conducted at 287 stations along the global coastline for the period 1994–2021.

Approximately 84.3% (242 stations) show strong correlation ($r \geq 0.7$) with TG for the NLT signal (Fig.4(a)), and 54.7% (157 stations) exhibit an RMSE≤20mm (Fig.5(a)). The NLT signal is influenced by decadal and/or multi-decadal oscillations, such as the El Niño-Southern Oscillation (ENSO) and the North Atlantic Oscillation (NAO), as well as long-term barystatic SL changes, steric changes, and variations in wind or ocean circulation. NLT signals generally exhibit a larger spatial extent. The effectiveness of DSR is particularly evident at stations along very narrow coasts like N. Spit in Humboldt Bay, located on NWA with a shelf-width of ~19km. Since DSR selects observations within a 250km SR, but only those within the shelf-break, observations as far as 247km from the TG (along the shelf) are selected and are found to capture the NLT signal observed at the coast by the TG.

Approximately 57% of the stations show good correlation with TG for the residual signal (Fig.4(b)), and only 21% exhibit a low RMSE (Fig.5(b)). Residual signals arise from various sources of monthly to interannual SL variability, such as those caused by CTW[63,64]. The satellite observations can capture residual signals in coastal regions of Western Australia, NWA, Baltic Sea, and the North Sea. A clear contrast is evident between the coasts of Eastern and Western Australia. Along the east coast, the EAC generates numerous highly variable mesoscale eddies, which can interfere with CTW signals near the shelf edge by influencing their energy flux and propagation characteristics[65,66]. Stations situated in tidally-dominated estuaries with high tides and turbidity, such as—Bay of Fundy, Rio de la Plata Estuary, and San Francisco Bay, show poor r and high RMSE. This poor residual signal correlation may arise from tidal influences often extending much farther upstream than saltwater intrusion in these estuaries, which in turn affects the ability of altimetry data to accurately capture the residual signal[67]. Stratification plays a significant role in the St. Lawrence Bay, which again can influence the propagation characteristics of internal waves[68] and hence we see poor r and RMSE.

About 98% (281 stations) show a strong correlation between altimetry and TG at an annual scale (Fig.4(c)). However, only 67% of the stations exhibit an RMSE ≤ 20mm (Fig.5(c)). Variations in the annual signal are primarily driven by factors such as wind stress, sea surface temperature, and salinity changes[69]. The Rimouski and Sept Isles stations located in the St. Lawrence Bay, the Port Chicago station in Suisun Bay (northeastern extension of San Francisco Bay), the Anchorage station in the Cook Inlet, and the Palermo station in Rio de la Plata show poor correlations with the TG data—all these stations are situated in highly turbid zones.

Finally, we assess the bias between TG and altimetry-derived SLA using stations referenced to the ellipsoid. A Student's *t*-test is applied to evaluate whether the bias is statistically significant. For most stations, we found no significant bias (p-value>0.05). The only exception is Rimouski, located at the mouth of the Rimouski River in Canada ($t$-statistic=2.43, p-value=0.01). The observed positive bias is likely attributable to an underestimation of the VLM rate at this station[70].

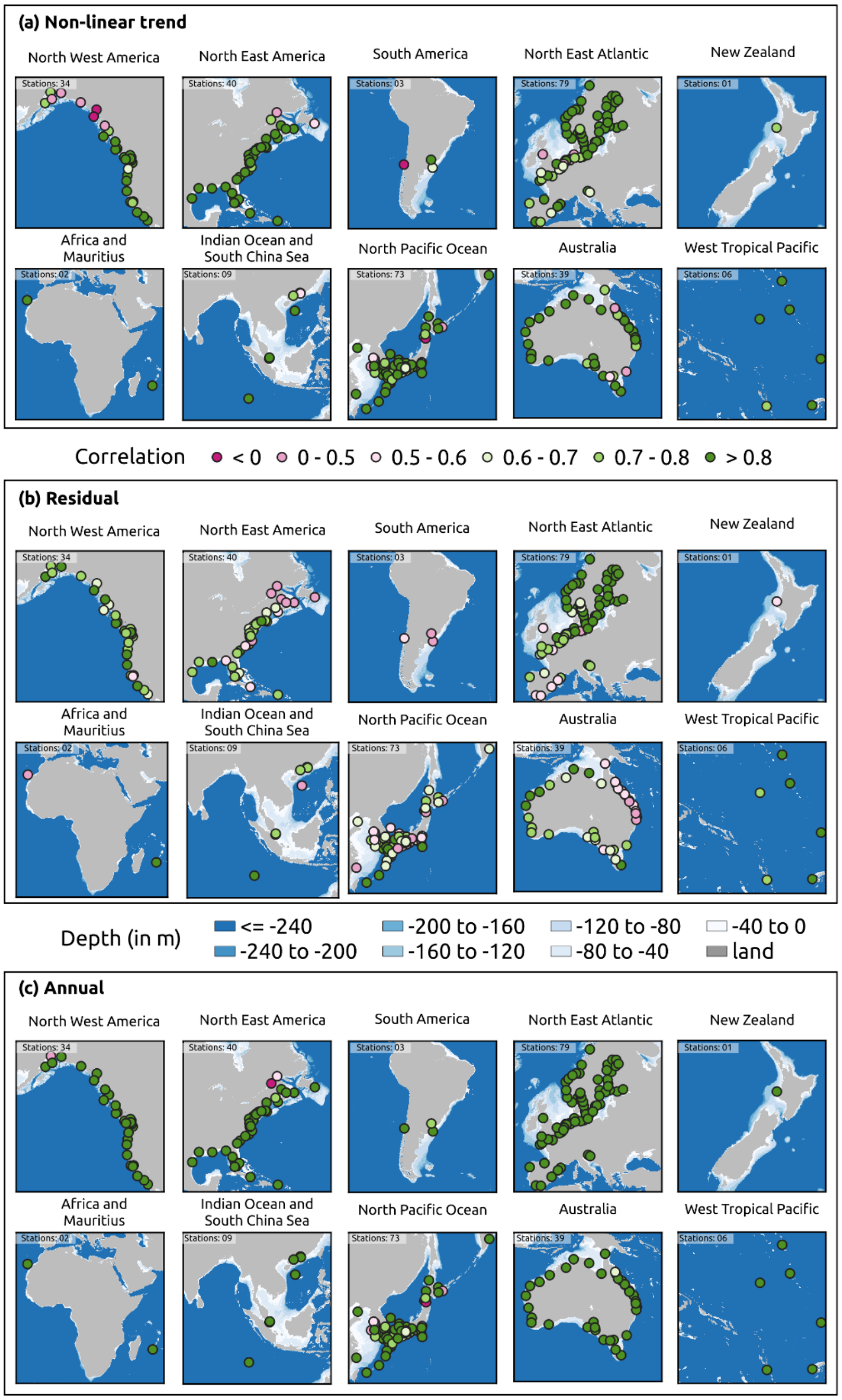

**Fig.4:** Assessment of XTRACK's performance in capturing SL variability across different temporal scales: Correlation of XTRACK SLA with TG for (a) NLT, (b) residual signal, and (c) annual signal. Green shades indicate a strong correlation (r), while pink shades represent a weak correlation.

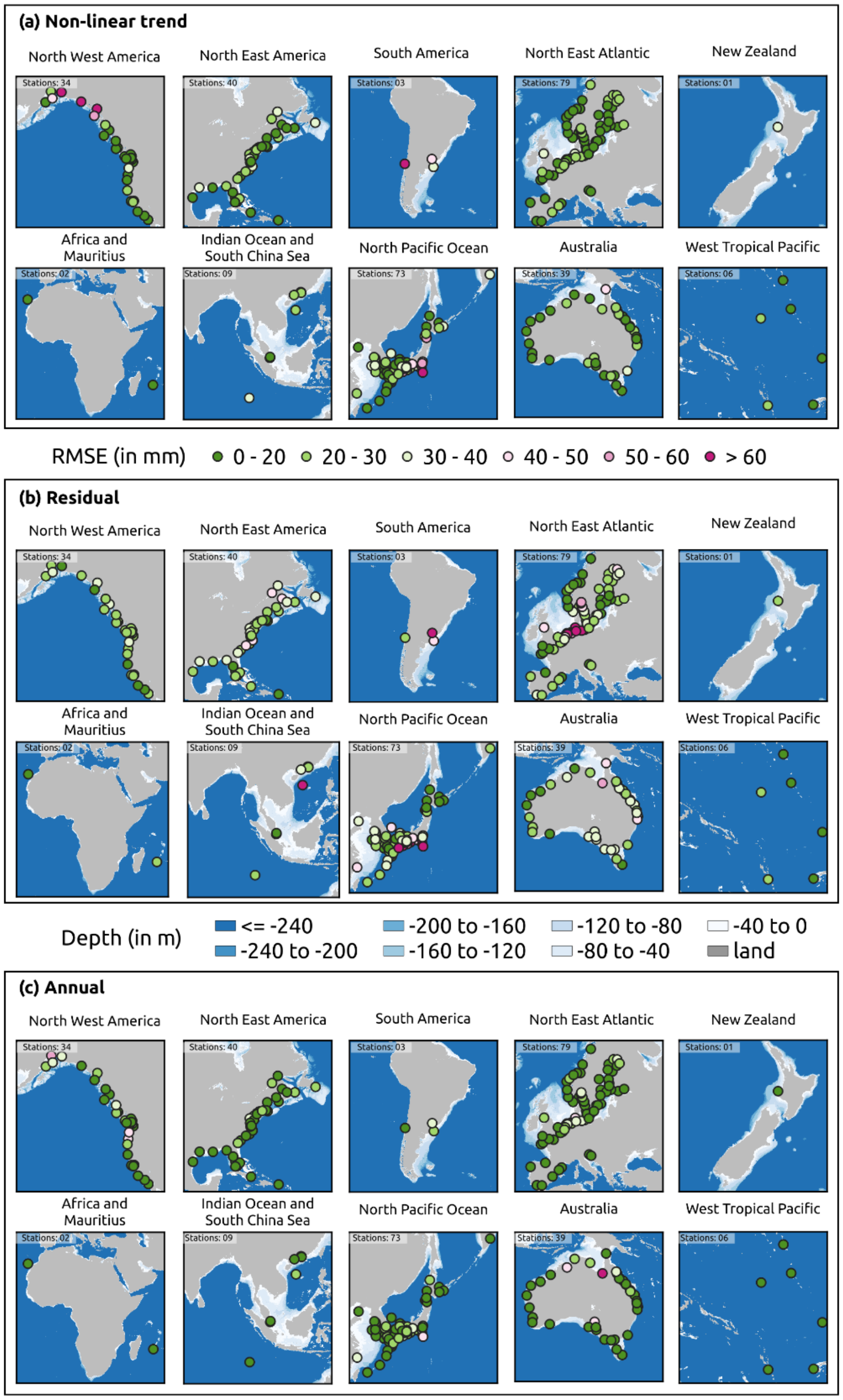

**Fig.5:** Assessment of XTRACK's performance in capturing SL variability across different temporal scales: RMSE of XTRACK SLA with TG for (a) NLT, (b) residual signal, and (c) annual signal. Green shades indicate low RMSE, while pink shades represent high RMSE.

## Coastal Sea Level Trends (CSLT)

CSLT over a 28-year period (1994–2021), is estimated at 267 stations (Fig.6). Trends estimated from TG are highest at Yokosuka, Japan with 10.2±0.6 mm/yr, while the altimetry-based trend is 5.12±0.83mm/yr. Here, persistent subsidence [71] may explain the ~10 mm/yr CSLT. The highest altimetry-based trend is observed at Onisaki, Japan (6.03±1mm/yr), where the TG-based trend is 5.81±0.87mm/yr.

The coastal SL rise is most pronounced in GoM, Baltic Sea, several stations in Japan, and low-latitude Northeast America (NEA). In the Baltic Sea, the mean trend is 4.6 ±1.4mm/yr from TG and 4.5±1.4mm/yr from altimetry. Similarly, GoM exhibits 4.5±1mm/yr from TG and 4.1±0.9mm/yr from altimetry. We also find the West Tropical Pacific (WTP) has a mean trend of 3.7±0.9mm/yr from TG and 4±0.9mm/yr from altimetry. Additionally, the North Sea coasts of Sweden and Norway show CSLT greater than the global mean SL trend of 3.4±0.4mm/yr, with 4±1 mm/yr from TG and 3.7±0.9 mm/yr from altimetry. In contrast, CSLT trends are much lower in NWA (1.6±1mm/yr from TG and 1.6±0.9mm/yr from altimetry) and in the Gulf of Alaska (1.3±1mm/yr from TG and 1.3±0.9mm/yr from altimetry). Earlier, Bonaduce et al. reported spatial variability in altimetry trends (mmyr$^{-1}$) across the Mediterranean Sea (1993–2012), with slightly higher values at locations such as Trieste (TG:3.19, altimetry:3.1) [72]. Using a longer TS, we find lower trends (TG:1.9 ± 0.6, altimetry:2.4 ± 0.6), suggesting that the higher values in the shorter record were possibly influenced by interannual variability.

In general, trends from altimetry (X_DSR) fall within the uncertainty limits of the TG estimates at about 71% of stations (189 stations) (CSLT trend uncertainties at 95% confidence level from XTRACK and TG are shown in Extended Data Fig.3). We find altimetry trends from X_DSR outside the uncertainty limits of TG trend estimates at 78 stations. Trends at 11 of these stations improved after TG stations were linked to the ellipsoid (see Methods section, Extended Data Fig.4-5). For example, the TG trend, referenced to Tide Gauge Benchmark (TGBM), at Stony Point is 0.82±0.78mm/yr and from X_DSR is 2.33±0.5mm/yr. After linking to the ellipsoid, the TG trend is 2.35±0.78mm/yr, closely matching the X_DSR trend. Unfortunately, only 16 of the 78 TG stations have ellipsoidal links. A full station list with trends from TG and XTRACK with uncertainties is given in Supplementary Table S2.

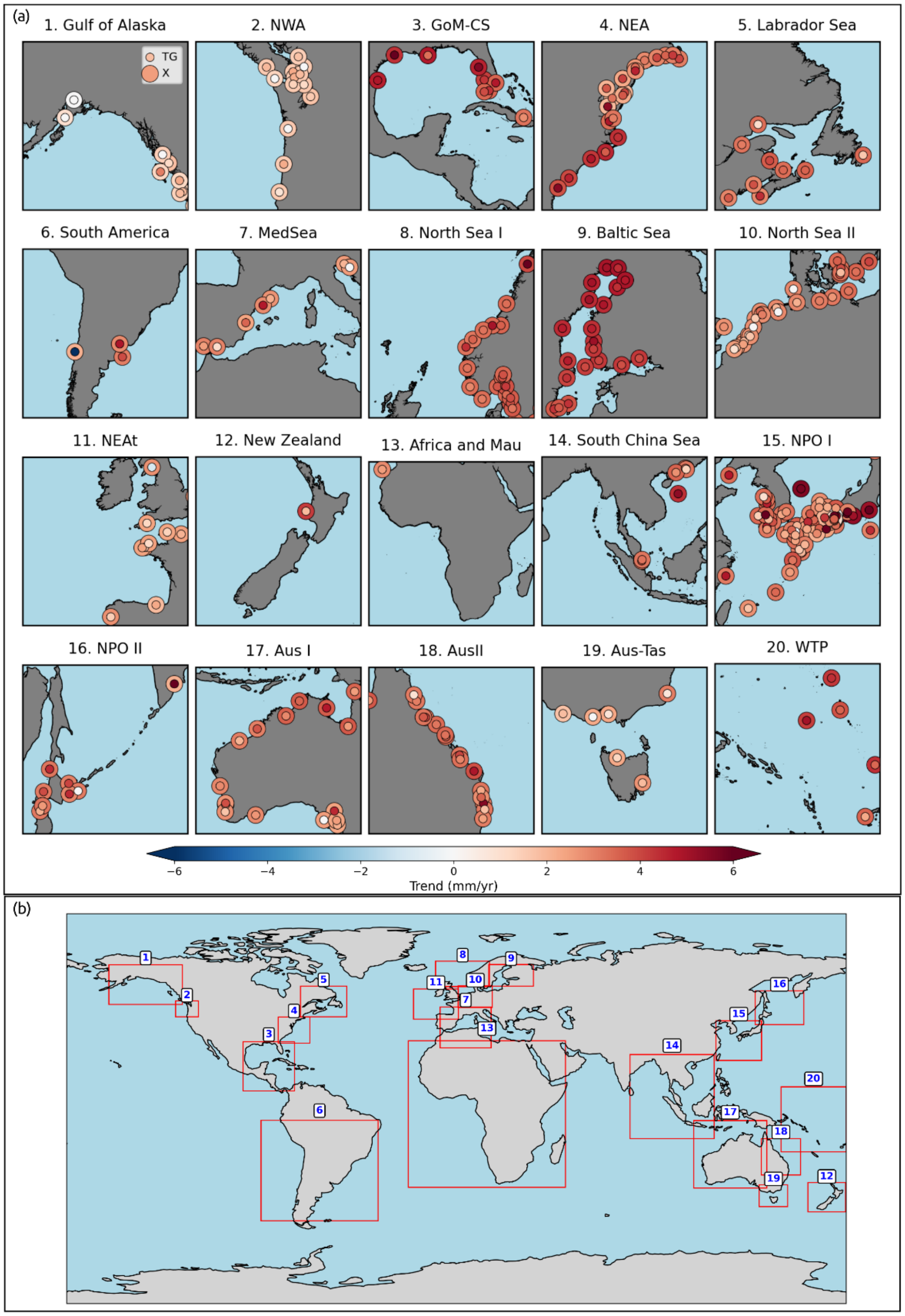

**Fig.6:** CSLT from 1994 to 2021 (28 years) derived from TG and coastal altimetry: Smaller circles represent trends from TG, and larger circles show trends from X_DSR. The colors indicate CSLT in mm/year, with red shades representing increasing trends and blue shades indicating decreasing trends. (a) Panels 1-20 shows region-wise SLA trends from TG and X_DSR and (b) shows world map and bounding boxes of region with box number corresponding to panel number.

## Conclusions

The coastal SL dynamics differ from the open ocean, primarily because ocean currents must flow parallel to the coastline[6]—the biggest influences being CTW and bottom effects[6]. Consequently, a novel altimetry data selection algorithm, called DSR is developed that chooses along-coast observations rather than open ocean points. The primary limitation of altimetry for coastal SL monitoring lies in its temporal and spatial sampling. Temporal sampling is improved by incorporating observations from multiple satellite passes, ensuring at least 15 per month. We found that restricting observations to within the continental shelf significantly improves the representation of coastal SL variability across multiple temporal scales, such as low-frequency—NLT (84% of stations have $r \geq 0.7$), seasonal—annual (98%), and monthly-to-interannual—residual (57%). The DSR algorithm consistently outperformed other altimetry selection algorithms, with the best performance observed across most regions, particularly in NEAt-NS, NPO, and SCS. We found that XTRACK/ALES provides limited improvement over XTRACK in capturing coastal SL variability; however, both struggled in tidally-dominated estuaries, St. Lawrence Bay, and along Eastern Australia. Trend estimates for over a period of 28-years showed altimetry trends well within the uncertainty limits of TG trends, with the highest CSLT observed in the GoM (4.1±0.9mm/yr), the Baltic Sea (4.5±1.4mm/yr) and several stations in Japan and low-latitude NEA. This work emphasizes that using DSR algorithm, altimetry is able to capture coastal SL variability effectively at monthly and longer temporal scales. Furthermore, the study demonstrates that the lower-resolution product (XTRACK) can be just as good as higher-resolution ones (XTRACK/ALES) in capturing SLA signals at these time scales. We also note that accounting for the influence of irregular and dramatically varying coastal morphology by region and within region can further improve the algorithm.

## Methods

### Data

Radar altimeter records the two–way travel time taken by radar pulses from satellite to sea surface and back, and is then converted to SSH using equation (1),

$$\text{SSH} = \text{orbital altitude} - (\text{range} + \text{corrections}) \quad (1)$$

Here, orbital altitude is the height at which the satellite flies with respect to the reference ellipsoid; range is the distance calculated from the time of return of the echo; corrections account for atmospheric effects on propagation time. We retain all standard geophysical corrections, including tidal and DAC, for comparison.

Closer to the coast, the return signal contains signatures from land that enter the altimeter footprint, the quasi-specular reflections from calm water (like in the docks), steeper waves, and short-scale variability of sea surface characteristics[12]. Yet, useful data can be retrieved by:

(1) Improving the range term by means of dedicated retracking algorithms for the coastal ocean during pre-processing (more effective for less than 10km to the coast)[12].

(2) Improving the corrections term during post-processing[12].

The two main sources of altimeter data in this study are:

1. X-TRACK SLA L2P Version 2022 (XTRACK): This product provides 1Hz SLA, which is equivalent to a distance between two consecutive along-track measurements of about 6-7 km. This product is AVISO L2P data (along-track sea level anomalies level-2+) post-processed using the X-TRACK program (described in Birol et al.[25]). It employs an efficient editing strategy to minimize the loss of observations near the coast and refines the correction terms specifically for coastal areas[25]. These corrections are statistically tested and recomputed using interpolation and extrapolation from datasets that provide local modelling of tides, atmospheric forcing, and wind stress[25].

2. Along-track sea level anomalies and trends v2.3 (2023) (XTRACK/ALES)[37]: This product provides 20Hz high-resolution monthly SLA dataset, which is equivalent to about 350m between two consecutive along-track measurements. This product is along-track altimetry preprocessed using the ALES retracking strategy[26] and post-processed using X-TRACK software[25]. The ALES retracking improves the range term and thereby the SSH estimates. It does so by discarding the trailing edge of the signal waveform where most of the contamination happens, and fitting sub-waveforms based on the detected SWH[26]. The sub-waveform width is adaptively adjusted to include the leading edge while minimizing contributions from the trailing edge, thereby enhancing the reliability of the range estimation[26]. For more details on ALES strategy, see Passaro et al.[22]. New network of virtual altimetry stations for measuring SL along the world coastlines v1, which provided data beyond 20km at 20Hz along track resolution, is considered obsolete and is no longer publicly accessible. Therefore, it was not used in this study.

TG data represent true monthly SLA mean values as it observes SL at hourly intervals. In contrast, satellite altimetry missions such as SARAL/AltiKa (35-day repeat cycle), Sentinel-3A (27-day repeat cycle), and the TOPEX/Poseidon + Jason series (10-day repeat cycle) collect data less frequently. This difference in temporal sampling can impact the ability of altimeters to capture SL level variations accurately. Birol et al.[36] highlighted that using only one observation per month can degrade the correlation between otherwise identical time series, reducing it from 1 to 0.7 on average. For these reasons, we chose the TOPEX/Poseidon and Jason series satellites, not only for their frequent observations (3-4 observations per month) but also for the long time series they provide, which is crucial for comparing SL variations at different temporal scales with TG data. Further, DSR has a threshold of 15 observations. This ensures the collection of observations from different passes and thus better sampling, and reduces the uncertainty associated with it. For instance, the altimetry SLA for the Brest station is produced using data from four satellite passes.

The Dynamic Atmospheric Correction (DAC) product is downloaded and is used to correct atmospheric perturbations on the TG SL observations. DAC data is available as a gridded product having a spatial resolution of 0.25° by 0.25° at 6-hourly temporal resolution. Monthly means are calculated using simple arithmetic means from the 6-hourly data at each grid point. For the DAC correction of TG SL, we considered the nine closest grid points and selected the one that provided the highest variance reduction in the DAC corrected SL. Estimates of VLM rates from GPS are also corrected at each PSMSL station. We flagged stations where VLM quality estimates are poor.

We did not include these stations for trend estimation. The VLM product used here is produced and distributed by the Nevada Geodetic Laboratory (NGL)[73,74].

GEBCO_2023 is a gridded global bathymetry dataset with a 15″ resolution and is used in this study to estimate the shelf-break along the global coastline. Preprocessing was carried out in QGIS, where the 200 m isobath was extracted using the "Contour" tool. Contours with an area smaller than 250 km² are filtered out and discarded (This was not applied to minor islands). Further, contours with an area between 250 $km^2$ and 400,000 $km^2$ that appeared spurious are also discarded (This was not applied on any islands). This exercise was employed to remove artifacts in the GEBCO bathymetry grids.

The in-situ TG data is obtained from the Permanent Service for Mean Sea Level (PSMSL). The monthly Revised Local Reference (RLR) SL data is used. In addition, ellipsoidal height links available for some stations from various GNSS solutions, also provided by PSMSL, are used.

**Station Selection Criteria**

Majorly, the study spans two time periods, 1994 to 2021 and 2002 to 2020. The former is in evaluating XTRACK's performance in capturing SL variability at various temporal scales and estimating long-term trends. The latter is for the intercomparison of XTRACK, XTRACK/ALES, and TG, where XTRACK SLA is computed using 7 different algorithms and XTRACK/ALES is derived using 5 alternative algorithms. The selection criteria for intercomparison with XTRACK/ALES and TG are,

1. TG observations available from 2002.

   This criterion aligns with the start of XTRACK/ALES data availability, which begins in 2002 with the Jason-1 satellite mission.

2. Complete data coverage for 19 years (January 2002–December 2020)

   This is to fix the study period to ensure consistency when comparing across different algorithms.

3. Data gap should not be more than 20% for the study period considered.

   This is to ensure continuity in SLA TS as gaps are interpolated using a simple linear model during deseasonalizing and detrending.

4. Only stations with satellite observations available within 100km of the TG from both coastal altimetry products (XTRACK and XTRACK/ALES) are chosen.

   We selected 100km because XTRACK/ALES uses more stringent editing, resulting in the exclusion of several satellite observations, and we aimed to include only those stations having observations within a 100km radius.

5. Any station with a continuous data gap of 4 months or more in the TS should be excluded.

   This is also to ensure that SLA TS does not have a continuous data gap.

After applying these criteria, a total of 155 TG stations are left for intercomparison. The XTRACK/ALES and XTRACK-derived SLA are compared with those of TG stations along the global coastline.

Additionally, we assessed the effectiveness of XTRACK in capturing SL variability across different temporal scales (low frequency, seasonal, and high frequency) by comparing its results with TG measurements at 287 stations. The selection criteria for evaluating XTRACK's performance using the DSR algorithm are,

1. TG observations are available from 1994: Although XTRACK data is available from 1993 with the TOPEX/Poseidon mission, the number of TG stations with 29 years of continuous data was less.
2. Complete data coverage for 28 years (January 1994 – December 2021): This is to fix the study period across all stations.
3. Data gap should not be more than 20% for the study period considered. This is to ensure continuity in SLA TS as gaps are interpolated using a simple linear model during TSD using STL and in trend estimation.
4. Only stations with satellite observations available within 350km of the TG are selected.
5. Any station with a continuous gap of 4 months or more in the TS should be excluded. This is also to ensure that TS does not have a continuous data gap.
6. There should be at least 15 satellite observations available for a station. This criterion has been explained in the section "The Dynamic Search Radius Algorithm".

**Statistical Metrics, Time Series Decomposition, and Trend Estimation**

For all comparisons, we required at least 80% completeness in the time series, and comparisons were made only when this condition was met[36,58,59]. For intercomparison with XTRACK/ALES and TG, we removed the annual and semi-annual signals from TG and XTRACK TS using a simple least-squares fit to sinusoidal function. Then the trend was computed using ordinary least squares (OLS) simple regression fit and removed. We used statistical metrics, Pearson correlation coefficient (r), Root Mean Square Error (RMSE), and Normalized Root Mean Square Error (NRMSE). We normalized RMSE using the standard deviation (sd) of TG SLA. The value of the NRMSE represents the ratio between the variation not explained by the altimetry SLA and the overall variation in TG SLA. If the altimetry SLA TS explains a fraction of the variation but leaves out some part, which is at a similar scale to the overall variation, the ratio will be around 1. If NRMSE is greater than 1, this indicates greater variation or noise in the altimetry SLA. For evaluating DSR algorithm performance, we did a TSD using the Seasonal and Trend decomposition using LOESS (STL) and analyzed NLT, residual, and annual signals. STL is chosen because it is resilient to transient outliers and allows annual amplitudes and trends to vary over time. We validated with TG by measuring r and RMSE at each of these temporal scales. After removing stations that are not corrected for VLM, the linear CSLT for the period of 1994 to 2021 (28 years) are estimated using ordinary least squares linear regression at 267 stations along the global coastline.

**Ellipsoidal Correction**

We analyzed the performance of DSR with TG linked to the ellipsoid. SL measured by the TG is with respect to a local datum, whereas altimetry measures sea surface height (SSH) above the reference ellipsoid, WGS84. Since we use RLR data from PSMSL, the stability of the local datum is ensured by fixing its height to a TGBM, which is assumed to be on stable ground. In our analysis, we correct for VLM using GPS data and then remove the long-term mean from SL to obtain the SLA, which is then compared with altimetry-derived SLA. By doing this, we assume the effect of data referenced to two different references will be small or negligible. For further analysis, we selected stations with information on ellipsoidal links from PSMSL (https://psmsl.org/data/obtaining/ellipsoidal_links.php). Ellipsoidal links are estimated using continuous GNSS measurements from different GPS solutions such as ULR7a, JPL14, NGL14, and GT3. All these solutions use GRS80 as the reference ellipsoid. We acquire 46 stations referred to the ellipsoid for comparison with X_DSR across different temporal scales. We apply ellipsoid correction as,

$$SSH_{elip} = SSH_{RLR} + H_{RLR\ datum} \quad (2)$$

Here, SSHelip is the SSH above the ellipsoid, $SSH_{RLR}$ is the SSH referred to the RLR TG datum, and $H_{RLR\ datum}$ is the height of the RLR datum above the ellipsoid. Additionally, this SSH is corrected for VLM using GPS solutions. This correction is necessary because, in equation (2), the RLR datum is assumed to be stable, which may not be true due to VLM. Further, we estimated bias between SLA from TG referenced to the ellipsoid and altimetry using Student's *t*-test.

**Sensitivity and Uncertainty Analysis**

To find the optimal parameter values to be employed in the algorithm, we carried out sensitivity analysis. In the DSR algorithm, the three parameters can be tweaked: (1) coastal extent, (2) threshold, and (3) SR2. These three parameters determine which altimetry observations are chosen. To evaluate how these factors influence data selection and, hence, the resulting altimetry SLA TS, we tried out 396 parameter combinations across 318 TG stations for the period 1994-2021. The combinations covered coastal extent from 5 to 50km in 5km increments, threshold values of 1 and from 5 to 25 in steps of 5, and SR2 from 150 to 350km in 50km increments. After applying the DSR algorithm, we performed TSD and computed r and RMSE between the altimetry and TG for NLT, seasonal, and residual signals. The first filtering step removed parameter combinations that discarded more than 20 stations, reducing the set from 396 to 198 combinations. For each remaining combination, we calculated the percentage of stations with $r^2 \geq 0.64$ (explaining at least 64% of variance), and the percentage of stations with RMSE ≤ 20 mm for NLT, residual, and annual signals. Based on the RMSE of NLT signals (Supplementary Fig.S1(b)), thresholds of 1 and 25 were eliminated, coastline extent of 5km was removed. Based on the $r^2$ of NLT signal (Supplementary Fig.S1(a)), the threshold of 20 was removed, SR2 values of 200km and 300km were removed, and coastline extent of 45km and 50km were removed. Based on the RMSE of the annual signal (Supplementary Fig.S3(b)), coastline extents of 45km and 50km were removed. Based on the $r^2$ of the annual signal (Supplementary Fig.S3(a)), the threshold of 5 and SR2 of 150km were removed.

After these filtering steps, 38 combinations remained. Among them, the combination of threshold as 15, coastline extent of 15km, and SR2 of 250km performed best based on RMSE and $r^2$ of residual (Supplementary Fig.S2(b) and Supplementary Fig.S2(a)), and thus are the selected parameters of DSR.

Additionally, to test the robustness of the algorithm, uncertainty in r at each station was estimated using the Fisher Z-transformation[75] (Supplementary Fig.S4).

**Acknowledgements**

VS acknowledges the Ministry of Education for the PhD fellowship. The authors acknowledge financial support from Anusandhan National Research Foundation (ANRF) (previously known as Science and Engineering Research Board (SERB)) under the grant agreement SRG/2022/000625. The authors express their gratitude to Diljit Dutta for discussions and his insightful comments.

**Data availability**

All data used are open source and are available at: in-situ tide gauge data from PSMSL (https://psmsl.org/), X-TRACK SLA L2P Version 2022 (referred to as XTRACK) from AVISO website (https://www.aviso.altimetry.fr/en/data/products/sea-surface-height-products/regional/x-track-sla/x-track-l2p-sla-version-2022.html) and Along-Track Sea Level Anomalies and Trends v2.3 (referred to as XTRACK/ALES) from SEANOE website (https://www.seanoe.org/data/00631/74354/), Dynamic Atmospheric Correction (DAC) data distributed freely by AVISO (https://www.aviso.altimetry.fr/en/data/products/auxiliary-products/dynamic-atmospheric-correction.html), Vertical Land Motion (VLM) rates at PSMSL stations from GPS network produced and distributed freely by Nevada Geodetic Laboratory (NGL)) and, GEBCO_2023, a gridded global bathymetry dataset (https://www.gebco.net/data_and_products/gridded_bathymetry_data/).

**Code availability**

All codes to reproduce the work will be deposited on GitHub and will be shared.

## Extended Data

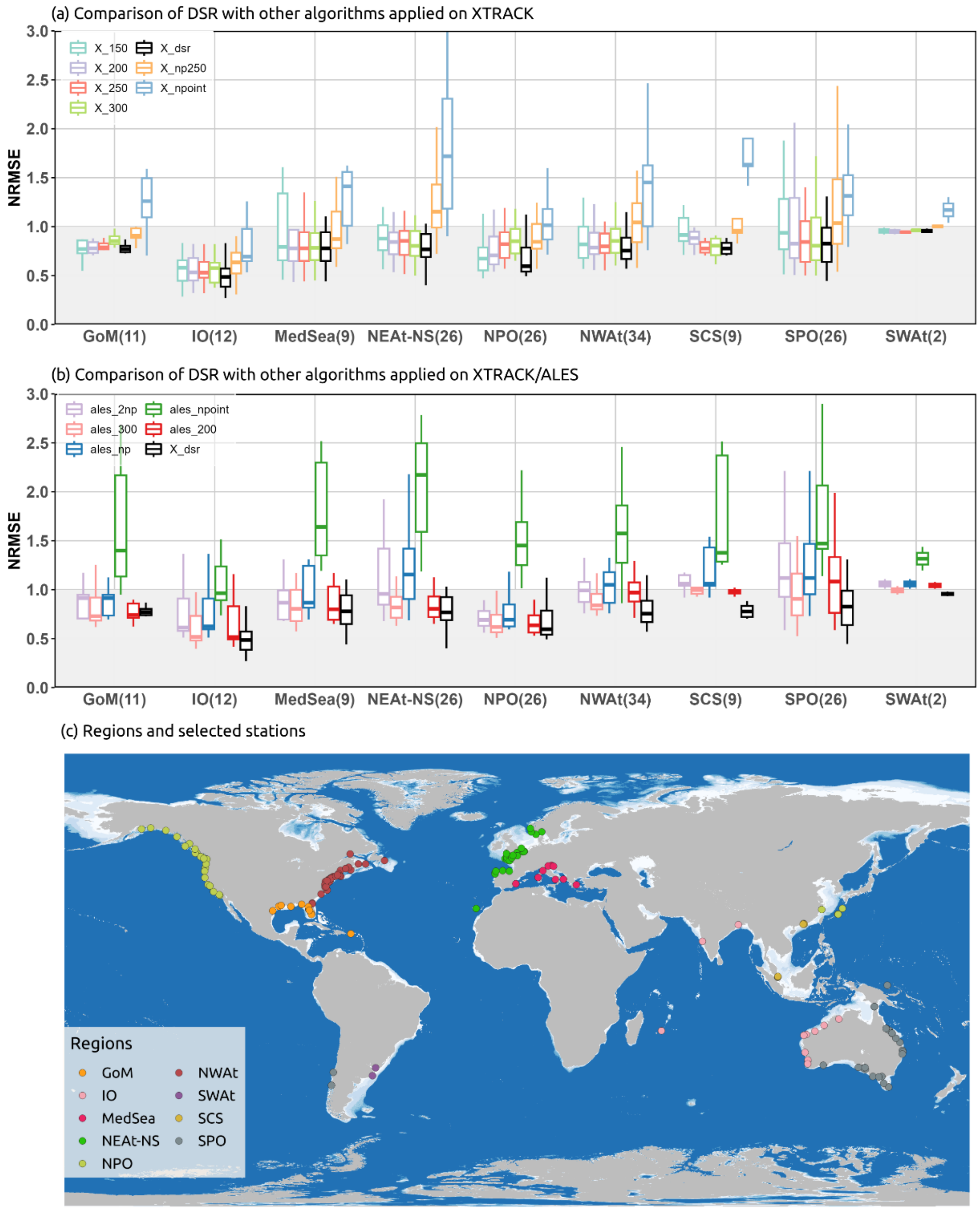

**Extended Data Fig.1:** Evaluation of different algorithms for representing coastal SL variability. (a) shows DSR (X_DSR) for XTRACK product compared with six alternate algorithms applied on XTRACK: X_150, X_200, X_250, X_300, X_np250 and X_npoint; (b) shows DSR (X_DSR) compared with five alternate algorithms applied on XTRACK/ALES: ales_2np, ales_300, ales_np, ales_npoint, ales_200. The comparison is based on NRMSE and is conducted at 155 TG stations.

Box plots of region-wise NRMSE is shown, with NRMSE<1considered as good agreement, shaded in grey. Stations from different regions are shown as colored circles, with each color representing a specific region.

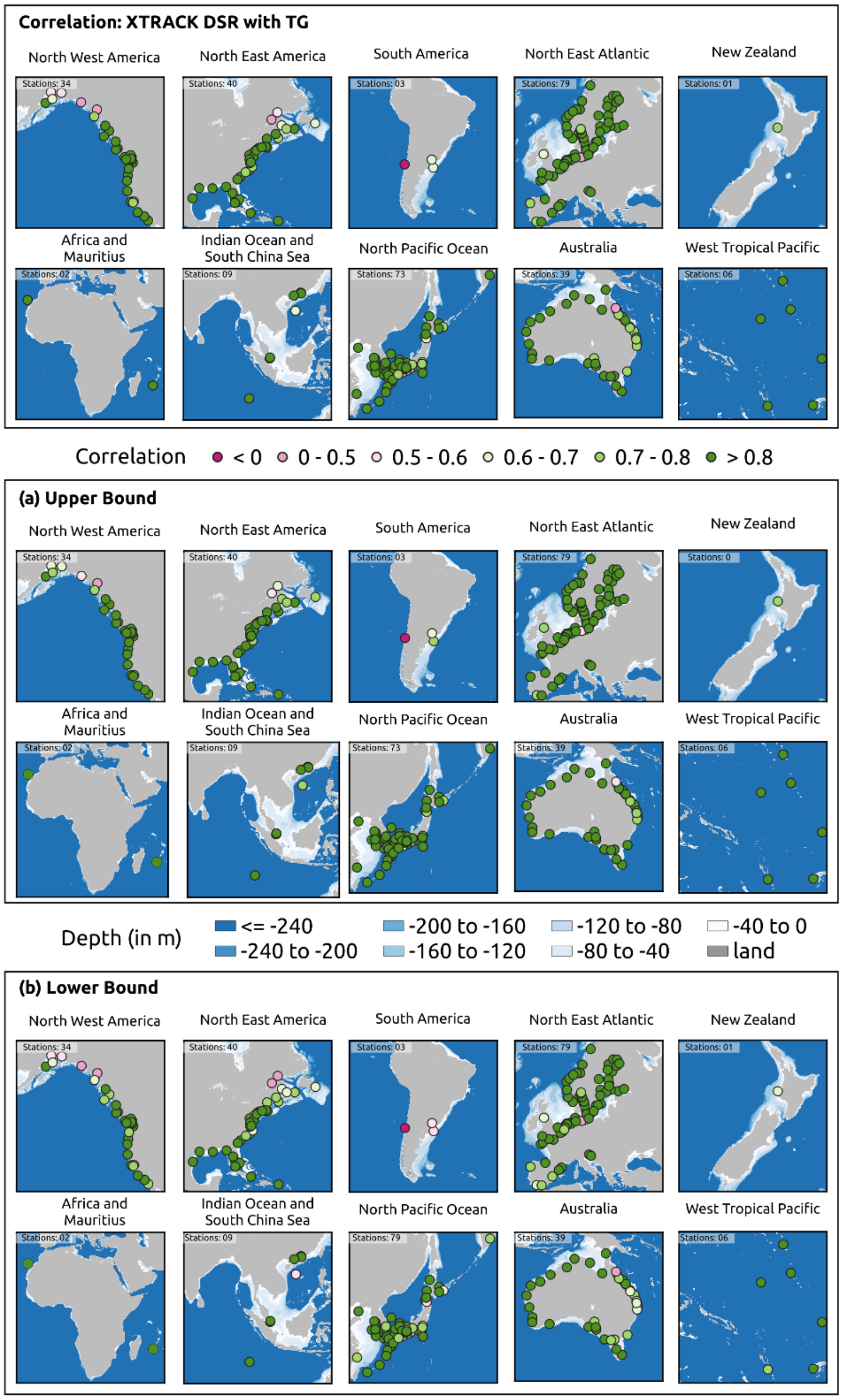

**Extended Data Fig.2:** Results from uncertainty in correlation coefficient at each station estimated using the Fisher Z-transformation. (a) Upper bound of correlation coefficient of SLA from X_DSR and TG, (b) lower bound of SLA from X_DSR and TG.

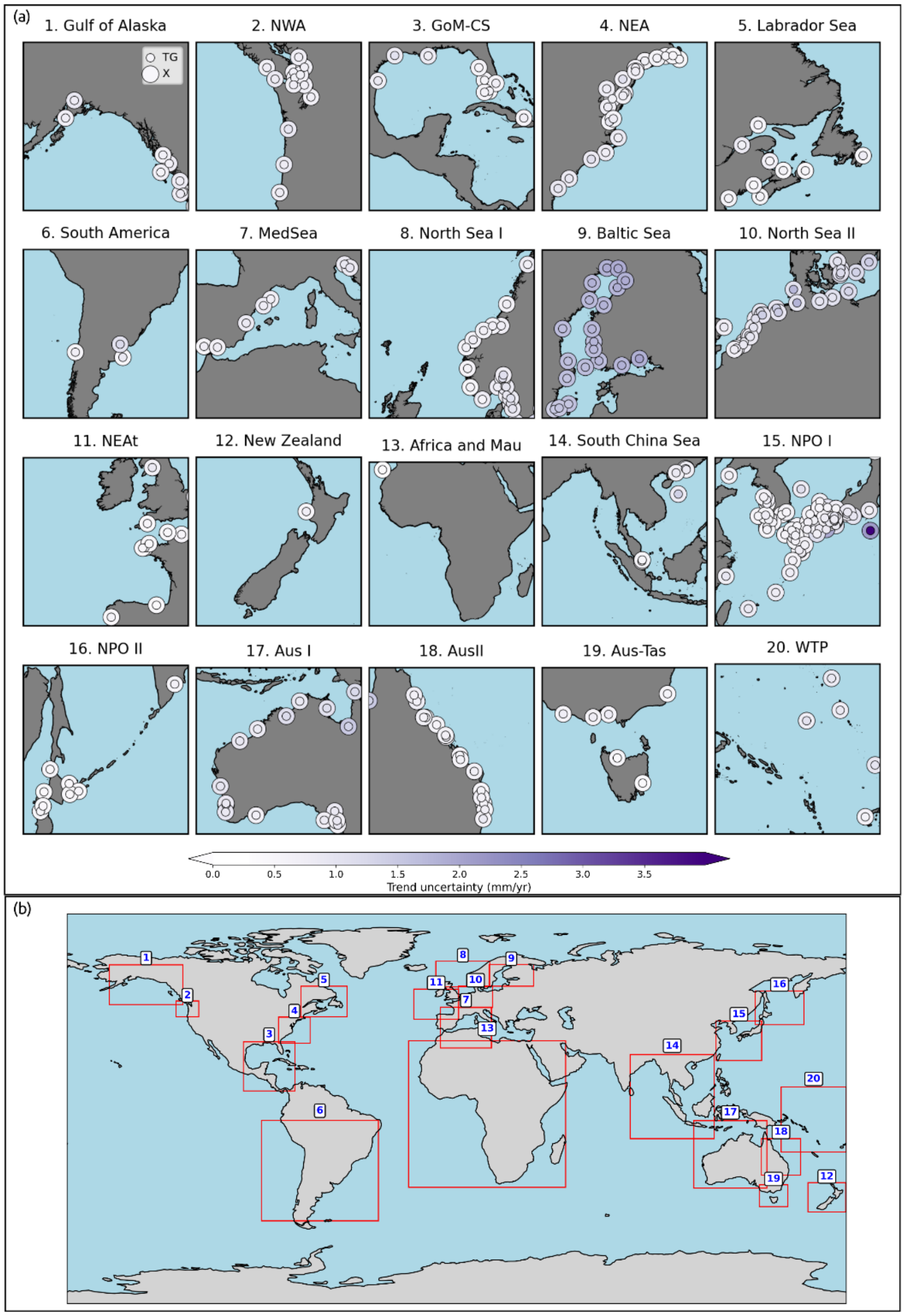

**Extended Data Fig.3:** Trend uncertainties of coastal SL from TG and altimetry (X_DSR) at 95% confidence level: Smaller circles represent trend uncertainties from TG, and larger circles show uncertainties from XTRACK (DSR algorithm). The colors indicate the magnitude of trend uncertainties in mm/year. (a) Panels 1-20 shows region-wise SLA trends from TG and X_DSR and (b) shows world map and bounding boxes of the regions with box number corresponding to panel number.

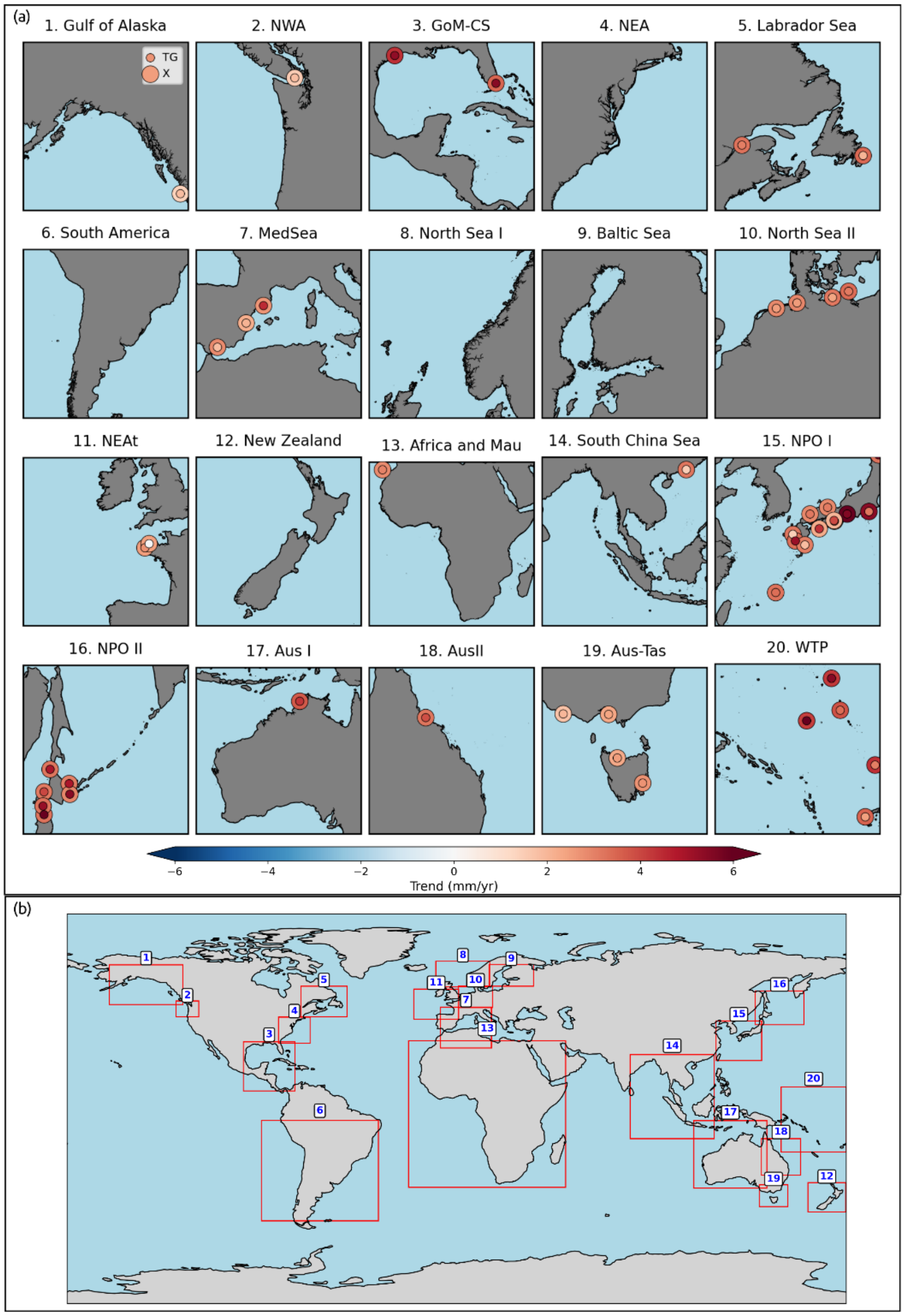

**Extended Data Fig.4:** CSLT from 1994 to 2021 (28 years) derived from TG referenced to ellipsoid and altimetry (XTRACK): Smaller circles represent trends from TG, and larger circles show trends from X_DSR. The colors indicate CSLT in mm/year, with red shades representing increasing trends and blue shades indicating decreasing trends. (a) Panels 1-20 shows region-wise SLA trends from TG and X_DSR and (b) shows world map and bounding boxes of the regions with box number corresponding to panel number.

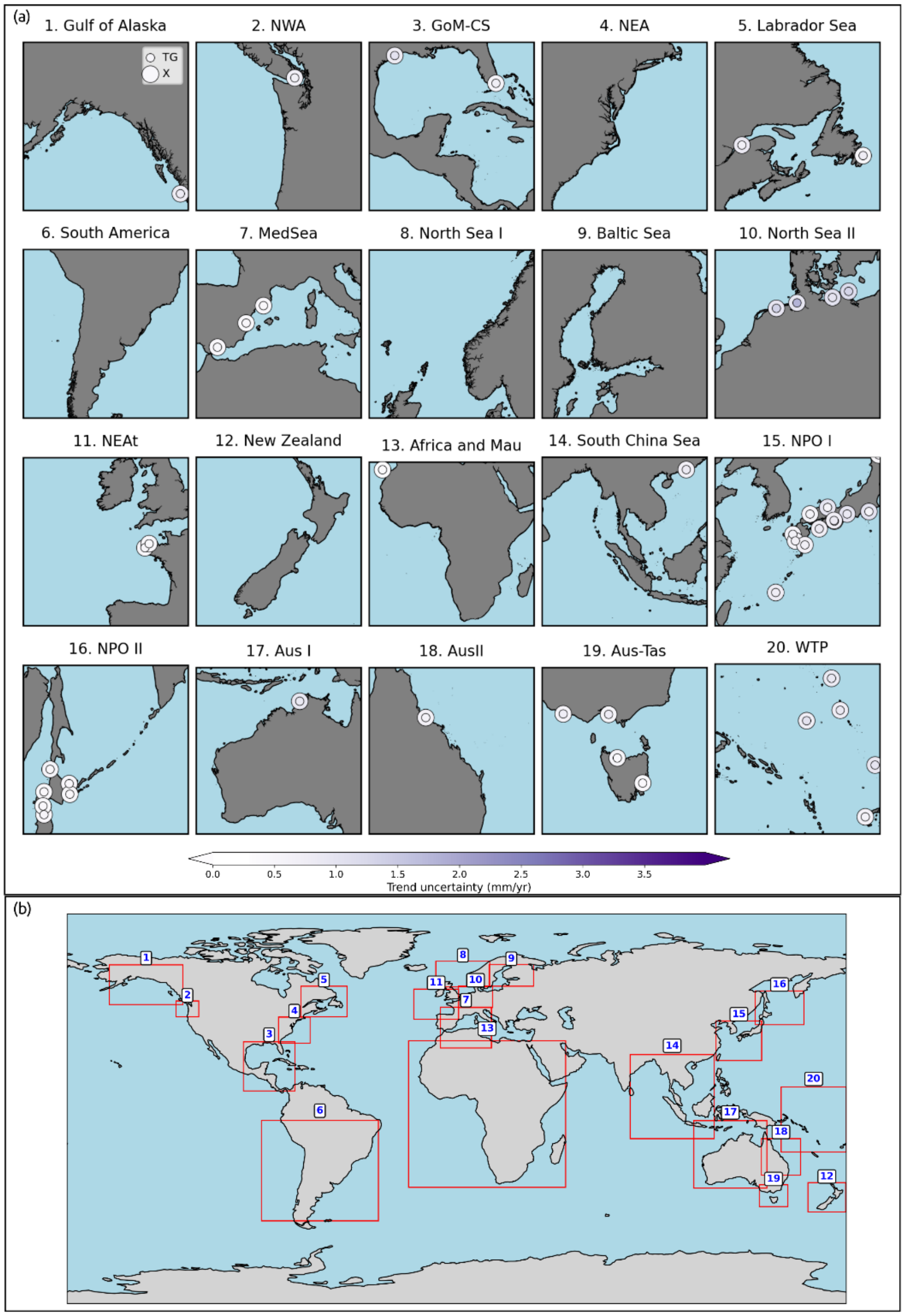

**Extended Data Fig.5:** Trend uncertainties of coastal SL from TG referenced to ellipsoid and altimetry (X_DSR) at 95% confidence level: Smaller circles represent trend uncertainties from TG and larger circles show uncertainties from XTRACK (DSR algorithm). The colors indicate the magnitude of trend uncertainties in mm/year. (a) Panels 1-20 shows region-wise SLA trends from TG and X_DSR and (b) shows world map and bounding boxes of the regions with box number corresponding to panel number.

Supporting Information for "Accounting for shelf width in selecting altimetry observations for coastal sea level variability improves its agreement with tide gauges"

Vandana S[1] and Bramha Dutt Vishwakarma[1,2]

[1]Interdisciplinary Centre for Water Research, Indian Institute of Science, Bangalore – 560012,

[2]Centre for Earth Sciences, Indian Institute of Science, Bangalore – 560012

Table of Contents

**Supplementary Text – Evaluating high frequency data (hourly)**

We used hourly tide gauge (TG) data from the University of Hawaii Sea Level Center (UHSLC) and applied preprocessing steps including tide removal, Dynamic Atmospheric Correction (DAC), and vertical land motion (VLM) correction. Tides were estimated from FES2022b tidal model. To ensure consistency with altimetry, we applied DAC from AVISO (6-hourly, linearly interpolated to hourly) using the nearest ocean grid point. VLM rates from nearby GPS stations were obtained from the Nevada Geodetic Laboratory (NGL) and applied as corrections. For comparison, we first identified the nearest altimetry point with at least 80% completeness in the SLA time series for 1994–2021, and then, for that altimetry pass containing that point, we selected the tide gauge observation closest in time to the satellite overpass. We then validated XTRACK-derived altimetry SLA against TG SLA. Supplementary Fig.S6 illustrates this comparison. Using statistical metrics, we obtained a correlation of 0.49 and an RMSE of 100.45mm.

## Supplementary Figures

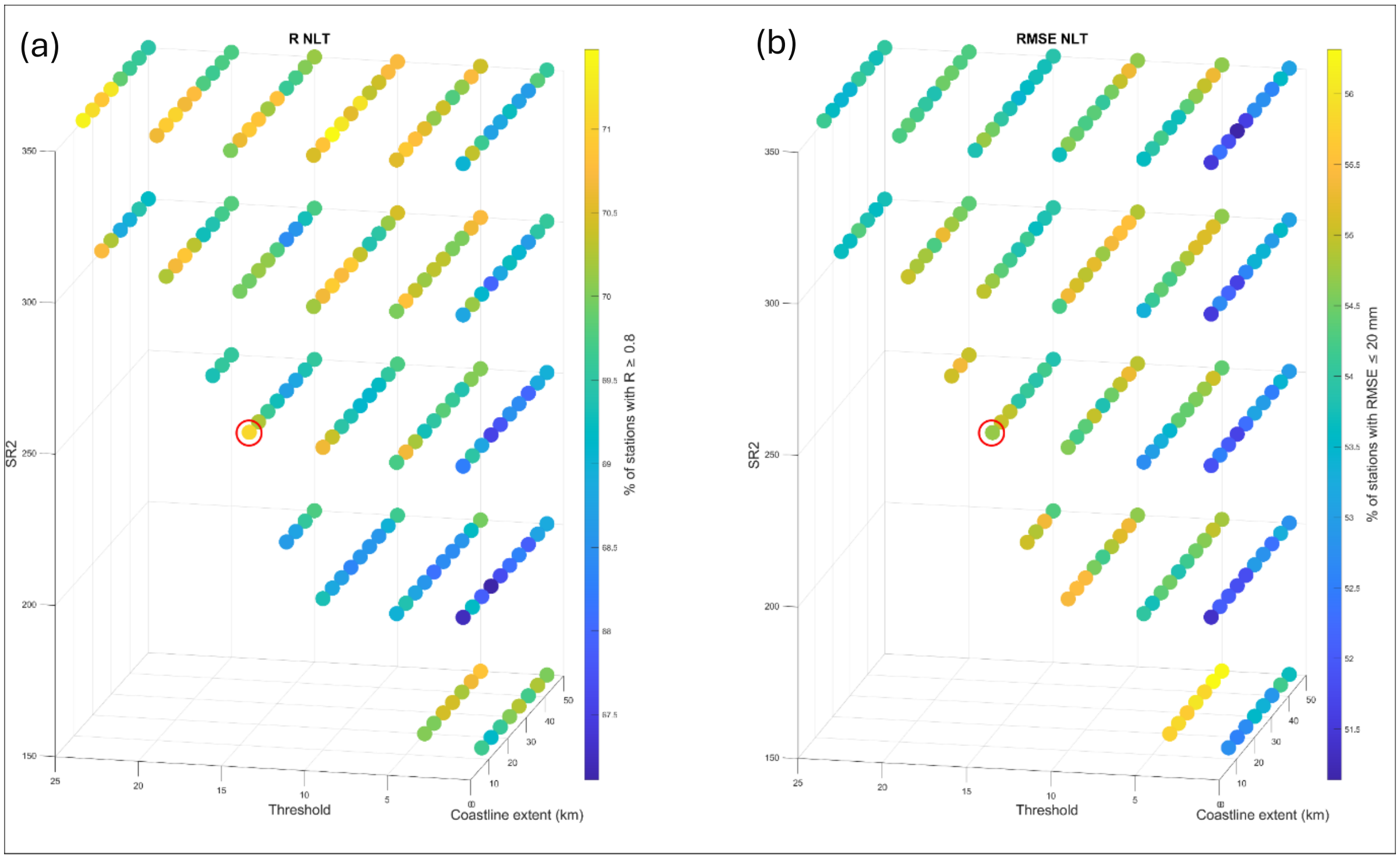

**Supplementary Figure S1. Three-dimensional scatter plots illustrating the sensitivity analysis across 198 parameter combinations of threshold, coastline extent, and SR2 (NLT).** The x-axis represents the threshold, the y-axis shows SR2 (in km), and the z-axis corresponds to coastline extent (in km). The colour of each circle in (a) denotes the percentage of stations with R ≥ 0.8 for Non-Linear Trend signal (R_NLT) and in (b) denotes the percentage of stations with RMSE ≤ 20 mm for Non-Linear Trend signal (RMSE_NLT). A red circle highlights the optimal parameter combination (15, 250, 15), identified as the best among 396 initial configurations. Only 198 combinations are shown, as combinations resulting in the exclusion of more than 20 out of 318 stations were removed.

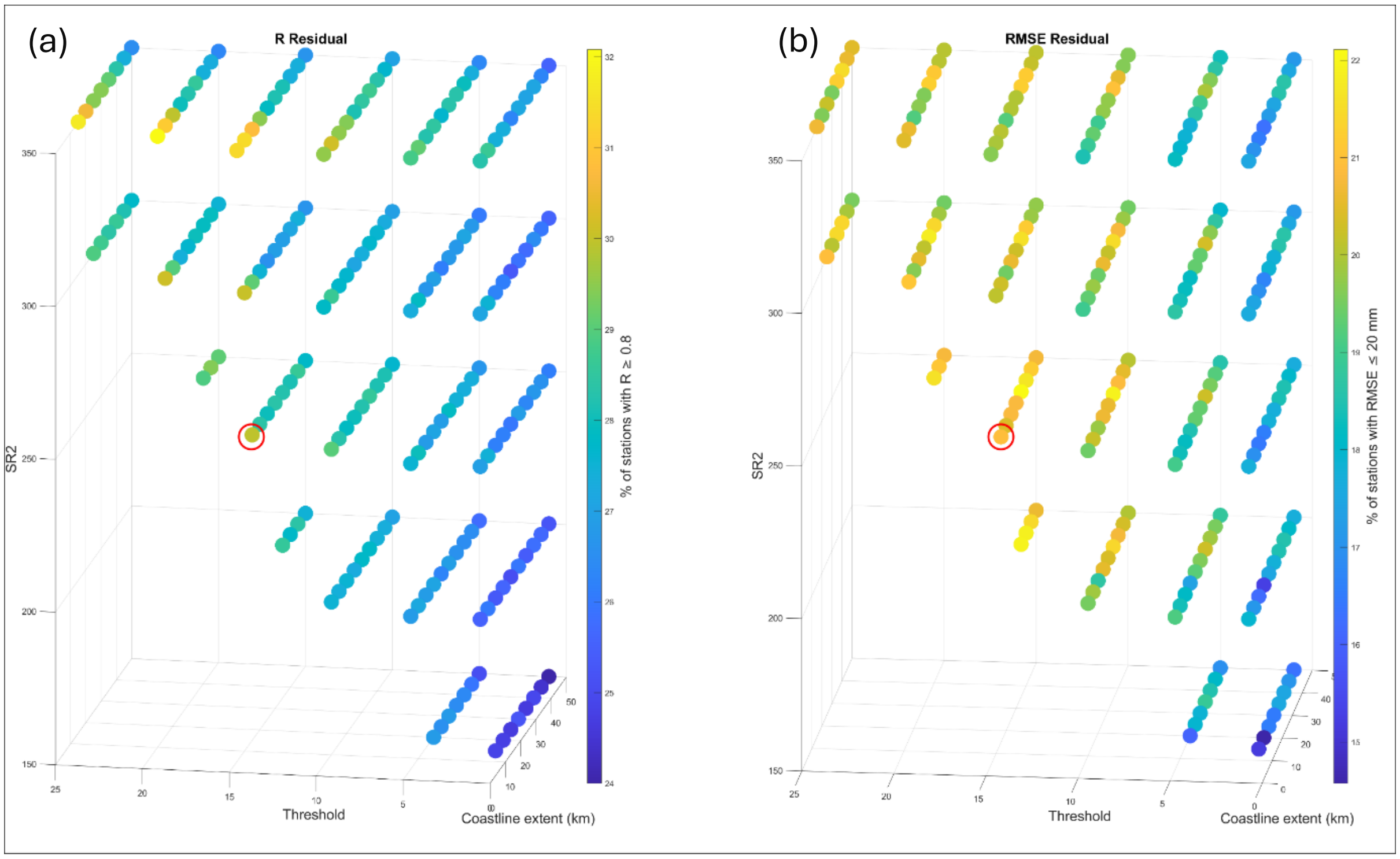

**Supplementary Figure S2. Three-dimensional scatter plots illustrating the sensitivity analysis across 198 parameter combinations of threshold, coastline extent, and SR2 (residual).** The x-axis represents the threshold, the y-axis shows SR2 (in km), and the z-axis corresponds to coastline extent (in km). The colour of each circle in (a) denotes the percentage of stations with R ≥ 0.8 for Residual signal (R_Residual) and in (b) denotes the percentage of stations with RMSE ≤ 20 mm for Residual signal (RMSE_Residual). A red circle highlights the optimal parameter combination (15, 250, 15), identified as the best among 396 initial configurations. Only 198 combinations are shown, as combinations resulting in the exclusion of more than 20 out of 318 stations were removed.

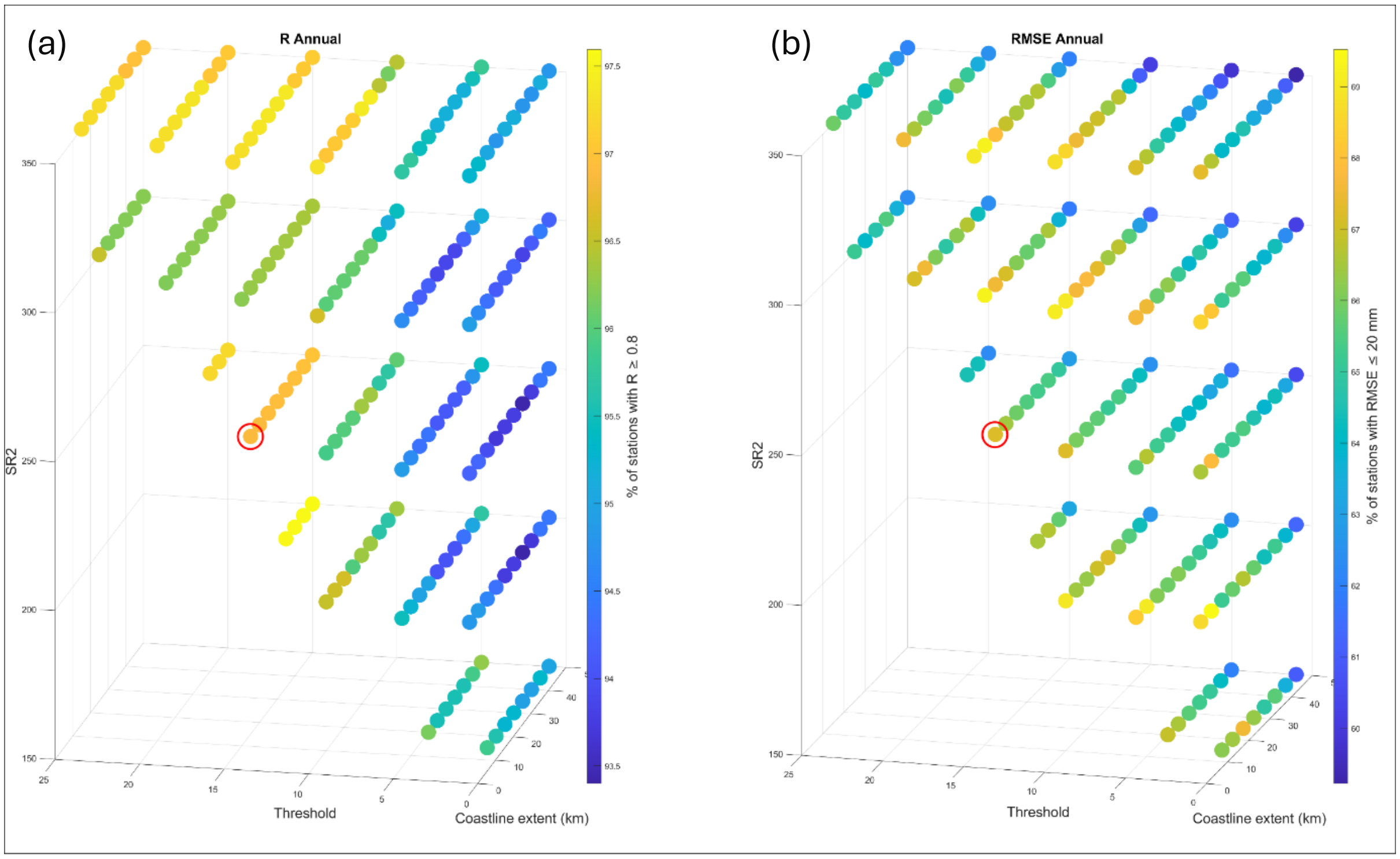

**Supplementary Figure S3. Three-dimensional scatter plots illustrating the sensitivity analysis across 198 parameter combinations of threshold, coastline extent, and SR2.** The x-axis represents the threshold, the y-axis shows SR2 (in km), and the z-axis corresponds to coastline extent (in km). The colour of each circle in (a) denotes the percentage of stations with R ≥ 0.8 for Annual signal (R_Annual) and in (b) denotes the percentage of stations with RMSE ≤ 20 mm for Annual signal (RMSE_Annual). A red circle highlights the optimal parameter combination (15, 250, 15), identified as the best among 396 initial configurations. Only 198 combinations are shown, as combinations resulting in the exclusion of more than 20 out of 318 stations were removed.

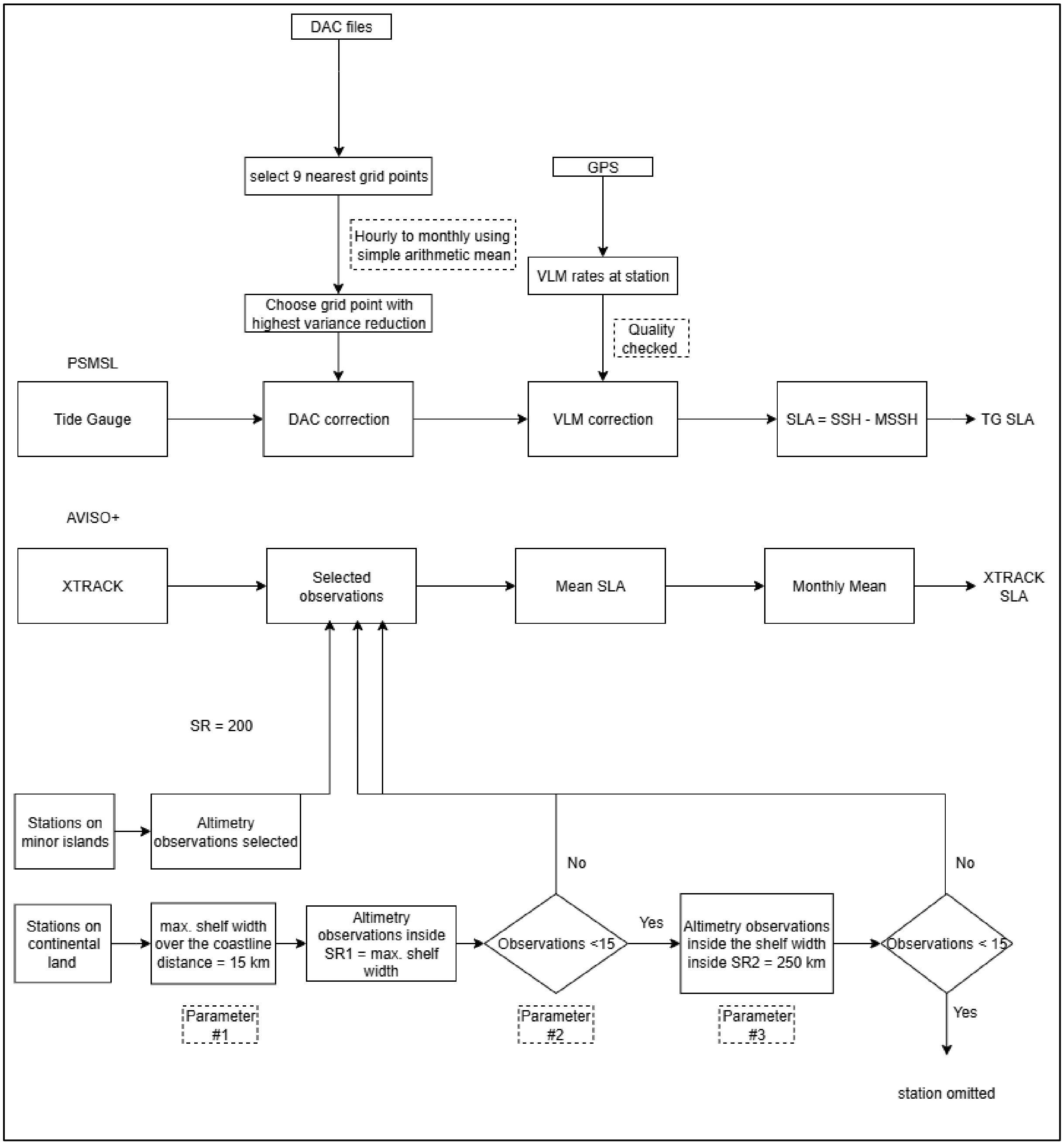

**Supplementary Figure S4. Flowchart for the selection algorithm for DSR.**

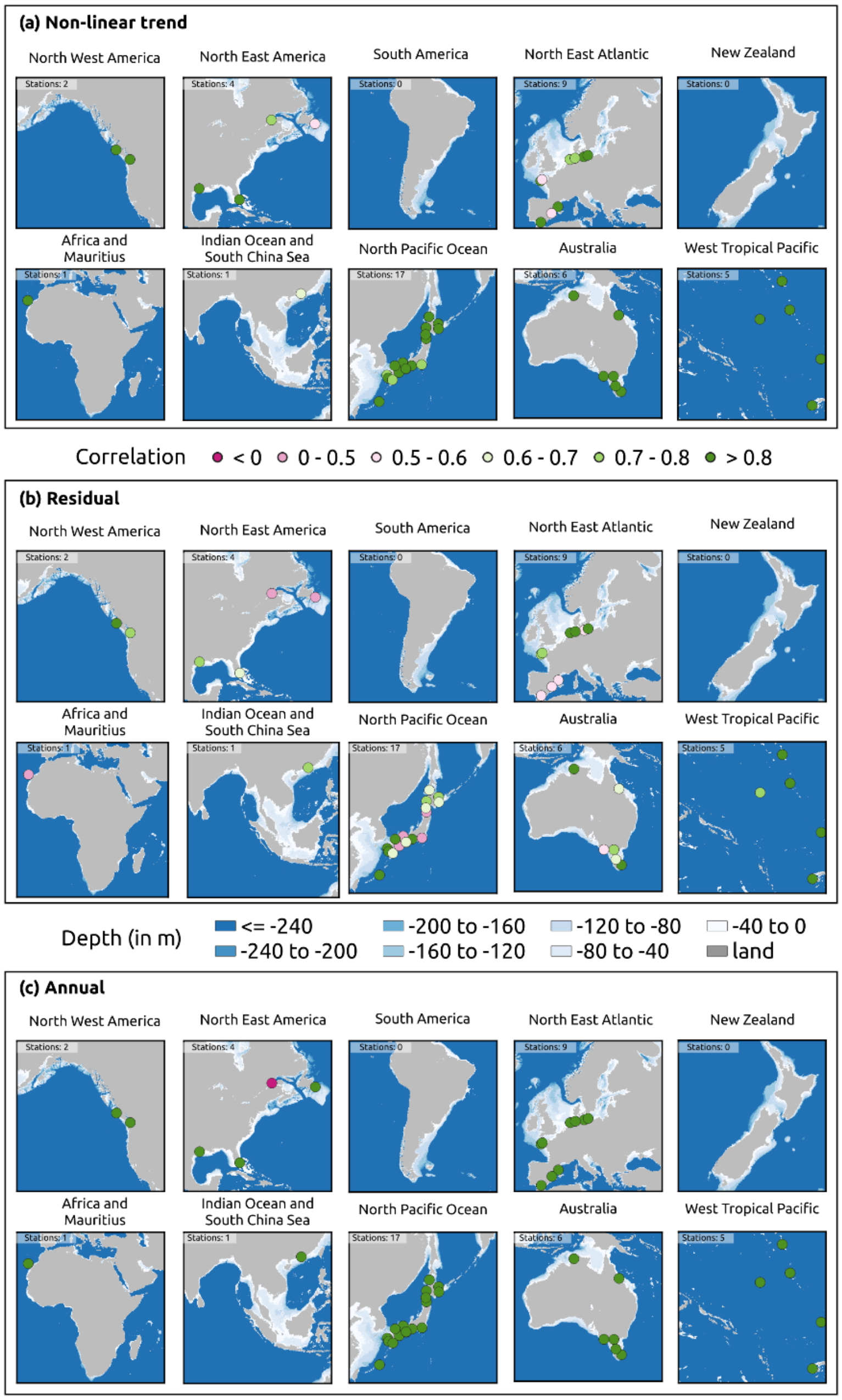

**Supplementary Figure S5. Assessment of XTRACK's performance in capturing sea level variability across different temporal scales with tide gauges (TG) linked to ellipsoid:** Correlation for (a) non-linear trend, (b) residual signal, and (c) annual signal. Green shades indicate a strong correlation (R), while pink shades represent a weak correlation.

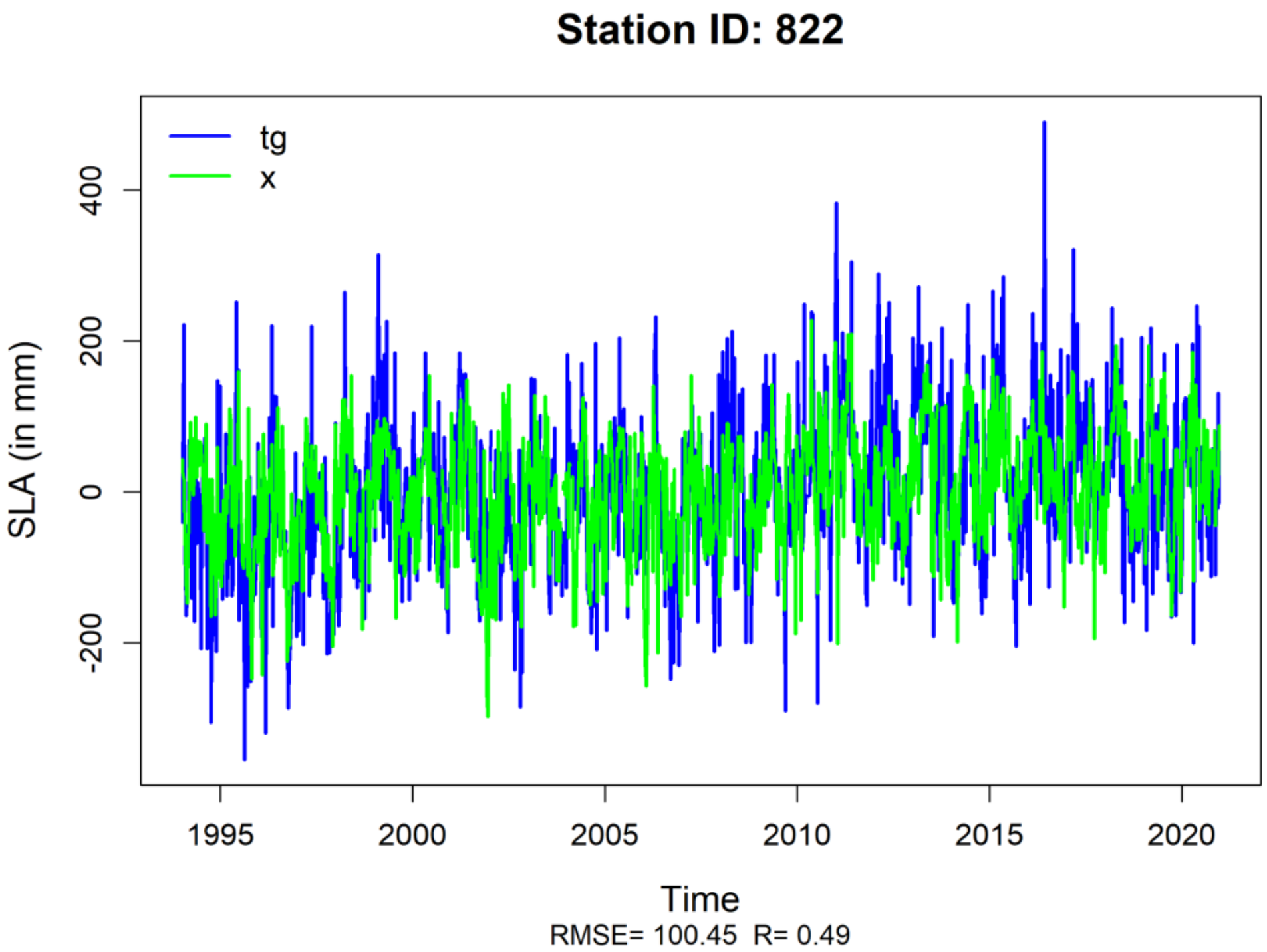

**Supplementary Figure S6. Comparison of high-frequency SLA (hourly) at Brisbane.** The figure shows the hourly time series of SLA from tide gauge (TG) data at Brisbane (from UHSLC) and from XTRACK. The TG data is corrected for DAC, tides, and VLM, and is compared with the nearest altimeter pass observation. The red line represents TG data, and the black line represents XTRACK data.

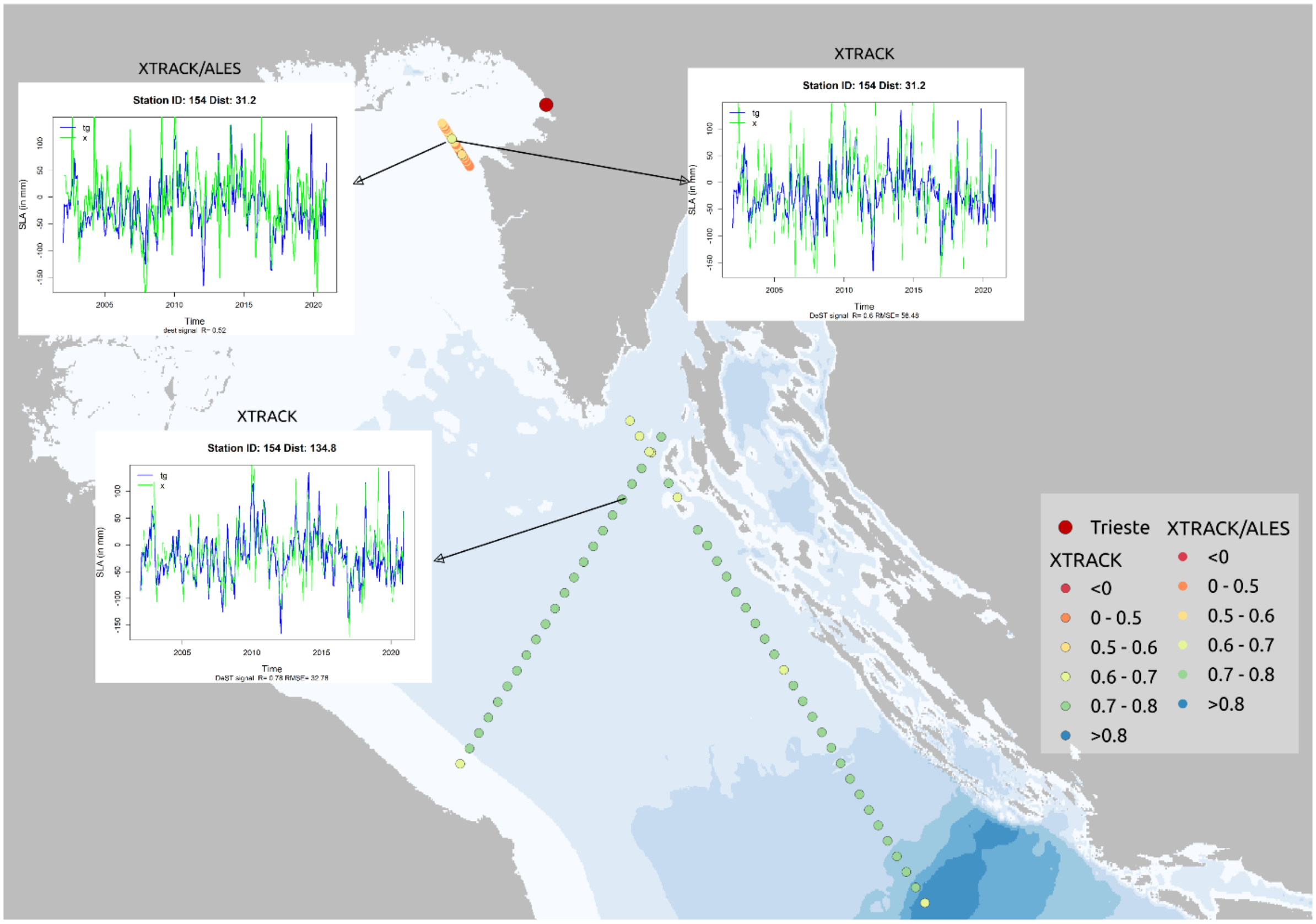

**Supplementary Figure S7. A deep dive into the capturing of sea level variability at Trieste.** The figure illustrates altimetry observations from XTRACK/ALES 20 Hz (circles without solid outlines) and XTRACK 1 Hz (circles with solid outlines). The observations show here are those selected using X_DSR algorithm and ales_300 algorithm. Colours represent correlation values between altimetry and tide gauge data, ranging from red (low, 0) to blue (high, 1). The arrows indicate the time series plot of altimetry sla at 31.2 km from XTRACK/ALES, 31.2 km from XTRACK and 134.8 km from XTRACK.

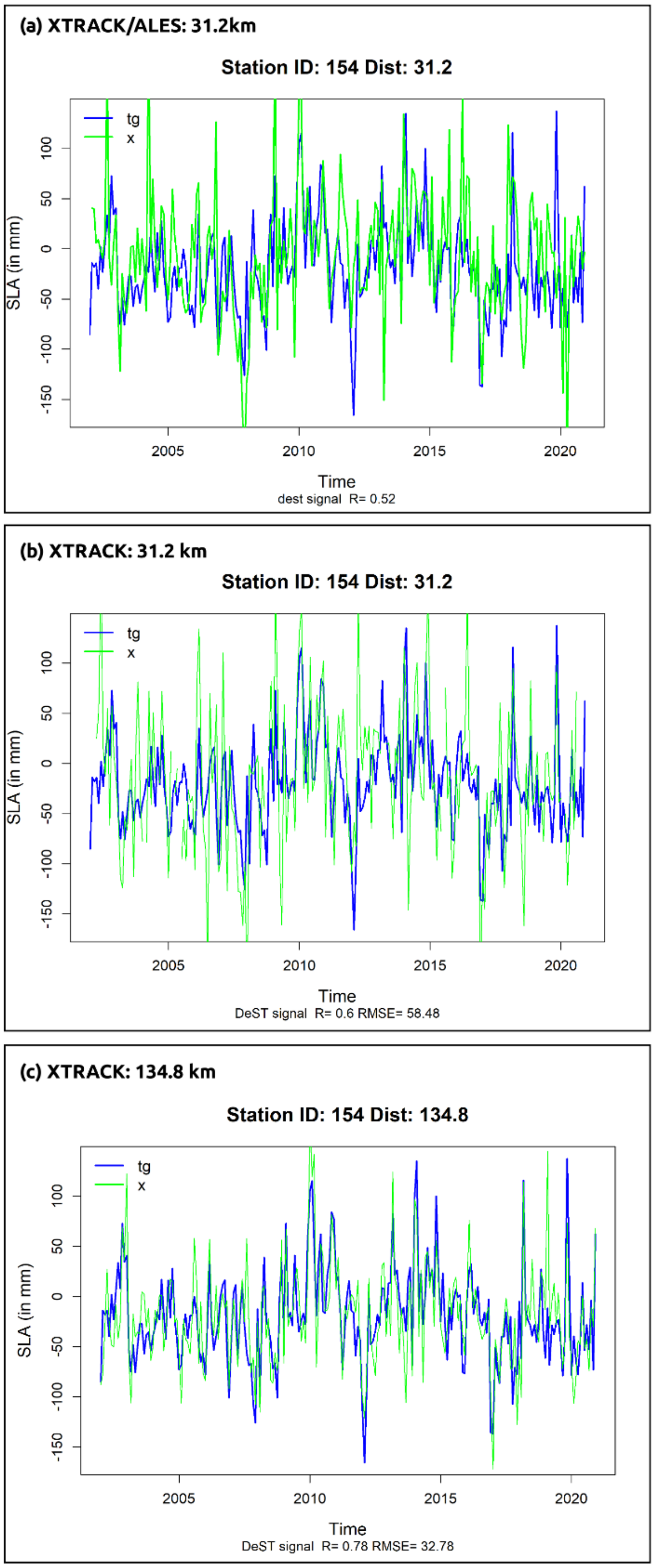

**Supplementary Figure S8. Time series comparison of coastal sea level variability (DeST) at Trieste.** The figure shows time series of sea level anomalies from tide gauge (TG; blue line) and altimetry (green line). (a) Comparison between TG and XTRACK/ALES observation at 31.2 km,

(b) Comparison between TG and XTRACK observation at 31.2 km, (c) Comparison between TG and XTRACK observation at 134.8 km

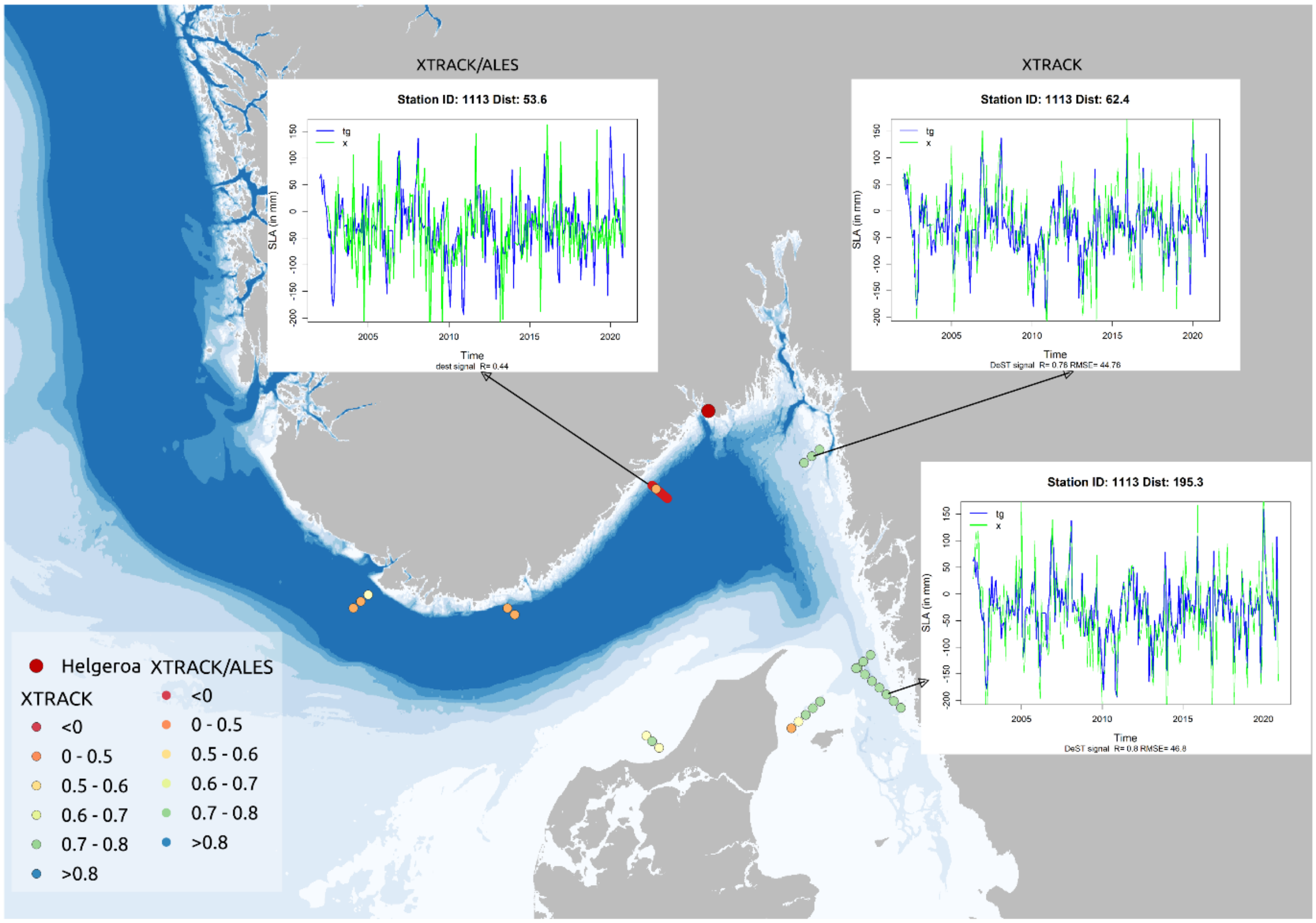

**Supplementary Figure S9. A deep dive into the capturing of sea level variability at Helgeroa.** The figure illustrates altimetry observations from XTRACK/ALES 20 Hz (circles without solid outlines) and XTRACK 1 Hz (circles with solid outlines). The observations show here are those selected using X_DSR algorithm and ales_300 algorithm. Colours represent correlation values between altimetry and tide gauge data, ranging from red (low, 0) to blue (high, 1). The arrows indicate the time series plot of altimetry sla at 53.6 km from XTRACK/ALES, 62.4 km from XTRACK and 195.3 km from XTRACK.

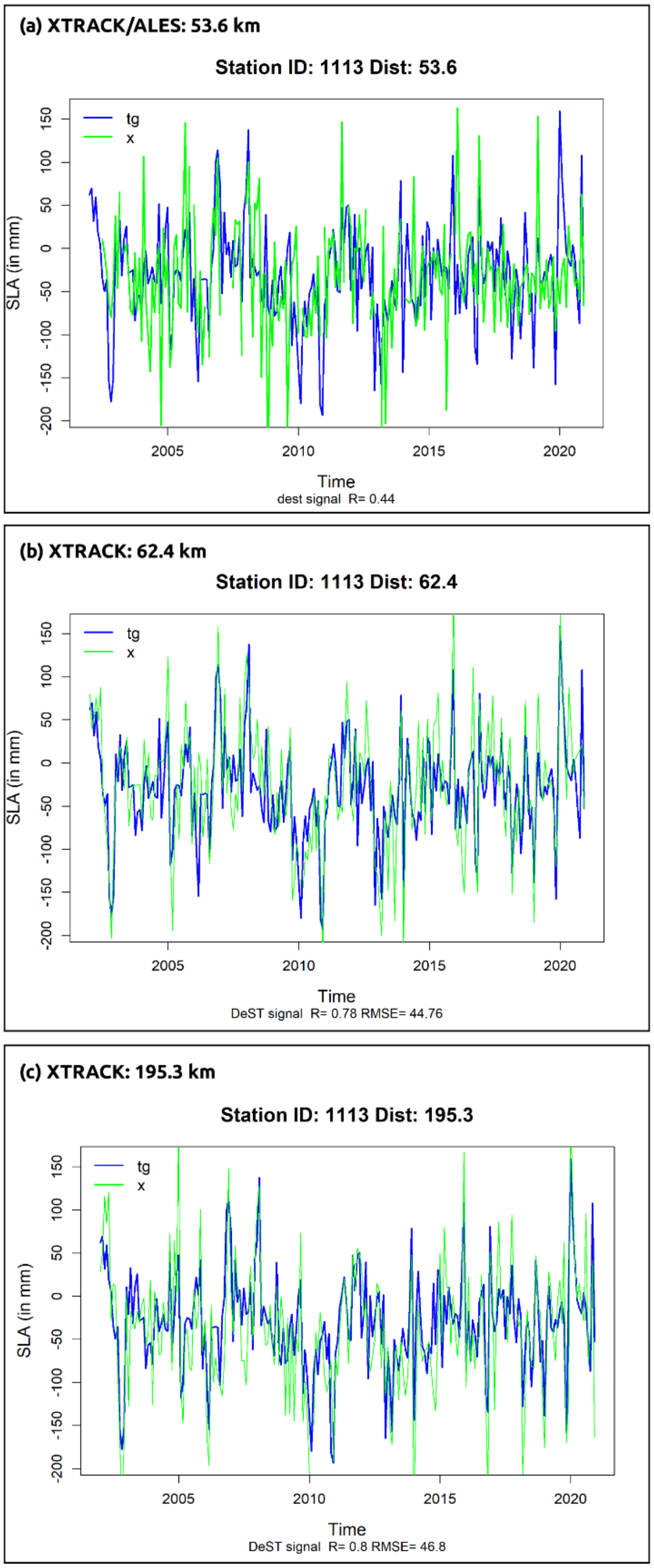

**Supplementary Figure S10. Time series comparison of coastal sea level variability (DeST) at Helgeroa**. The figure shows time series of sea level anomalies from tide gauge (TG; blue line) and altimetry (green line). (a) Comparison between TG and XTRACK/ALES observation at 53.6 km,

(b) Comparison between TG and XTRACK observation at 62.4 km, (c) Comparison between TG and XTRACK observation at 195.3 km

**Supporting Tables**

**Supplementary Table S1. List of stations on minor islands.**

| Station Name | Latitude | Longitude |
|---|---|---|
| Majuro-C | 7.106028 | 171.3728 |
| Lautoka | -17.6049 | 177.4383 |
| Betio | 1.365056 | 172.933 |
| Nauru-B | -0.52942 | 166.9101 |
| Honiara-B | -9.42892 | 159.9554 |
| Apra Harbor | 13.43833 | 144.6533 |
| Lombrum | -2.04208 | 147.3738 |
| Suva-B | -18.1325 | 178.4275 |
| Chichijima | 27.08333 | 142.1833 |
| Port Vila B | -17.7553 | 168.3077 |
| Funafuti B | -8.50247 | 179.1952 |
| Lord Howe Island | -31.5167 | 159.0667 |
| Xi Sha | 16.83333 | 112.3333 |
| Nase Iii | 28.5 | 129.5 |
| Ishigaki Ii | 24.33222 | 124.1636 |
| Okinawa | 26.17944 | 127.8244 |
| Nakano Sima | 29.84194 | 129.8478 |
| Naha | 26.21333 | 127.6653 |
| Kozu Sima | 34.20917 | 139.1317 |
| Miyake Sima | 34.06722 | 139.4808 |
| Okada | 34.78944 | 139.3914 |
| Kaminato Ii (Hatizyo Sima) | 33.13028 | 139.8047 |
| Ulleung | 37.49139 | 130.9136 |
| Port Louis Ii | -20.15 | 57.5 |
| Cocos Island (Home Is.) | -12.1167 | 96.89442 |
| Tenerife | 28.47722 | -16.2411 |
| San Juan | 18.45833 | -66.115 |
| Midway Island | 28.21167 | -177.36 |
| Kawaihae, Hawaii Island | 20.03667 | -155.83 |
| Rarotonga B | -21.2048 | -159.785 |
| Mokuoloe Island | 21.43167 | -157.79 |
| Kahului Harbor, Maui Island | 20.895 | -156.477 |
| Nuku'alofa B | -21.1368 | -175.181 |
| Hilo, Hawaii Island | 19.73 | -155.055 |
| Honolulu | 21.30667 | -157.867 |
| Nawiliwili Bay, Kauai Island | 21.95333 | -159.355 |

**Supplementary Table S2. Coastal sea level trend with uncertainty at 267 stations for the period 1994 to 2021 from tide gauges and altimetry (DSR).**

| Station Id | Station Name | Lat | Lon | TG trend | DSR trend |
|---|---|---|---|---|---|
| 1 | Brest | 48.38285 | -4.49484 | 1.8±0.48 | 2.51±0.43 |
| 10 | San Francisco | 37.80667 | -122.465 | 1.58±0.68 | 1.89±0.51 |
| 1001 | San Juan | 18.45833 | -66.115 | 2.84±0.5 | 3.13±0.47 |
| 1026 | Onisaki | 34.90389 | 136.8236 | 5.81±0.87 | 6.03±0.99 |
| 1027 | Oshoro Ii | 43.20944 | 140.8581 | 3.9±0.39 | 3.13±0.4 |
| 1028 | Hirosima | 34.35306 | 132.4647 | 3.09±0.68 | 2.82±0.57 |
| 1031 | Geraldton | -28.776 | 114.6019 | 2.86±0.94 | 2.91±0.93 |
| 1033 | Stony Point | -38.3721 | 145.2247 | 0.82±0.78 | 2.33±0.5 |
| 1034 | Tai Po Kau, Tolo Harbour | 22.4425 | 114.1839 | 2.79±0.88 | 3.2±0.66 |
| 1036 | Amrum (Wittduen) | 54.61667 | 8.383333 | 0±1.74 | 2.93±1.01 |
| 1037 | Borkum (Fischerbalje) | 53.5575 | 6.747778 | 2.01±1.36 | 2.88±0.93 |
| 1062 | Matsuyama Ii | 33.85889 | 132.7122 | 3.57±0.66 | 3.08±0.77 |
| 1066 | Jeju | 33.5275 | 126.5431 | 4.64±0.52 | 3.18±0.52 |
| 1067 | Anchorage | 61.23833 | -149.89 | 0±0.79 | 0±0.89 |
| 1068 | Bridgeport | 41.17333 | -73.1817 | 3.14±0.68 | 2.82±0.57 |
| 1069 | Victor Harbour | -35.5625 | 138.6354 | 2.2±0.88 | 2.34±0.68 |
| 1070 | Seldovia | 59.44 | -151.72 | 0±0.72 | 1.01±0.67 |
| 1071 | Port Hardy | 50.71667 | -127.483 | 1.89±0.69 | 1.32±0.69 |
| 1089 | Maisaka | 34.68194 | 137.6089 | 5.1±0.93 | 5.31±0.95 |
| 1094 | Hakata | 33.61889 | 130.4078 | 1.98±0.54 | 3.57±0.64 |
| 1095 | Saigo | 36.20139 | 133.3311 | 2.11±0.52 | 2.79±0.7 |
| 1096 | Nisinoomote | 30.735 | 130.9922 | 1.93±0.67 | 3.1±0.52 |
| 1097 | Odomari | 31.02361 | 130.6892 | 1.6±0.76 | 3.1±0.61 |
| 1098 | Sumoto | 34.34056 | 134.9067 | 2.22±0.81 | 0.91±0.71 |
| 1099 | Osaka | 34.65806 | 135.4328 | 5.45±0.79 | 0±0.76 |
| 11 | Warnemunde 2 | 54.16972 | 12.10333 | 2.66±0.98 | 3.16±1.05 |
| 1100 | Nagasaki | 32.735 | 129.8661 | 2.98±0.61 | 3.19±0.54 |
| 1101 | Fukue | 32.69611 | 128.8494 | 2.2±0.57 | 3.17±0.52 |
| 1102 | Oura | 32.97667 | 130.2206 | 0±0.64 | 3.16±0.59 |
| 1103 | Wakkanai | 45.40778 | 141.6853 | 4.5±0.4 | 3.21±0.55 |
| 1104 | Abashiri | 44.01944 | 144.2858 | 4.04±0.42 | 3±0.43 |
| 1105 | Nakano Sima | 29.84194 | 129.8478 | 3.2±0.65 | 3.8±0.57 |
| 1107 | Naples | 26.13167 | -81.8067 | 4.27±0.53 | 4.13±0.58 |
| 111 | Fremantle | -32.0558 | 115.7394 | 2.39±0.89 | 2.63±0.87 |
| 1111 | Nantucket Island | 41.285 | -70.0967 | 4.04±0.49 | 3.06±0.45 |
| 1113 | Helgeroa | 58.99521 | 9.856379 | 3.78±0.75 | 3.51±0.77 |
| 1114 | Esperance | -33.8709 | 121.8954 | 2.66±0.65 | 2.91±0.63 |
| 1116 | Wyndham | -15.4533 | 128.101 | 2.9±1.16 | 3.59±1.1 |
| 1147 | Sasebo Ii | 33.15806 | 129.7239 | 3.48±0.59 | 3.13±0.57 |
| 1148 | Tajiri | 35.59361 | 134.3158 | 2.23±0.57 | 2.87±0.96 |
| 1150 | Shirahama | 33.68361 | 135.3753 | 3.32±0.83 | 3.27±1.05 |
| 1151 | Naha | 26.21333 | 127.6653 | 3.43±0.76 | 3.28±0.8 |
| 1152 | Patricia Bay | 48.65 | -123.45 | 2.27±0.67 | 1.58±0.63 |
| 1153 | Cape May | 38.96833 | -74.96 | 3.98±0.71 | 2.1±0.65 |
| 1154 | Bundaberg, Burnett Heads | -24.7667 | 152.3833 | 4.92±0.65 | 3.86±0.6 |
| 1158 | Yarmouth | 43.83333 | -66.1333 | 4.43±0.58 | 2.87±0.51 |
| 1159 | Broome | -18.0008 | 122.2186 | 3.79±0.82 | 3.03±0.81 |
| 1160 | Milner Bay (Groote Eylandt) | -13.8601 | 136.4156 | 4.74±1.12 | 3.65±1.03 |
| 1196 | South Beach | 44.625 | -124.042 | 2.21±0.79 | 1.53±0.83 |
| 12 | New York (The Battery) | 40.7 | -74.0133 | 3.29±0.76 | 2.62±0.66 |
| 1209 | Kure Iv | 34.24056 | 132.5503 | 3.46±0.68 | 2.67±0.55 |
| 1210 | Tokuyama Ii | 34.04083 | 131.8028 | 4.47±0.66 | 2.77±0.56 |
| 1211 | Spikarna | 62.36333 | 17.53111 | 4.71±1.66 | 4.47±1.67 |
| 1212 | Uwajima Ii | 33.22722 | 132.5511 | 2.42±0.65 | 2.2±0.62 |
| 1216 | Spring Bay | -42.5459 | 147.9327 | 1.91±0.37 | 2.68±0.36 |
| 1236 | Goteborg - Torshamnen | 57.68472 | 11.79056 | 3.84±0.95 | 3.77±0.8 |
| 1241 | Rorvik | 64.85946 | 11.23011 | 3.37±0.73 | 3.46±0.59 |
| 1242 | Bamfield | 48.85 | -125.133 | 1.97±0.76 | 1.55±0.68 |
| 1246 | Hay Point | -21.2833 | 149.3 | 3.54±0.66 | 3.21±0.53 |
| 1265 | Akune | 32.0175 | 130.1908 | 2.42±0.62 | 3.33±0.55 |
| 1269 | Charleston Ii | 43.345 | -124.322 | 0.81±0.76 | 1.58±0.7 |
| 127 | Seattle | 47.60167 | -122.338 | 1.58±0.69 | 1.73±0.62 |
| 1293 | Oita Ii | 33.26611 | 131.6867 | 2.32±0.67 | 2.53±0.57 |
| 1294 | Le Conquet | 48.3594 | -4.78083 | 2.38±0.48 | 2.44±0.42 |
| 1295 | Cambridge Ii | 38.57333 | -76.0683 | 3.27±0.67 | 2.06±0.66 |
| 1299 | North Sydney | 46.21667 | -60.25 | 3.55±0.48 | 3.48±0.39 |
| 13 | Travemunde | 53.95806 | 10.87222 | 3.08±0.96 | 3.12±1.04 |
| 1320 | Kure I | 33.33361 | 133.2433 | 3.56±0.84 | 2.23±0.72 |
| 1322 | Sept-Iles | 50.2 | -66.4 | 1.06±0.64 | 3.37±0.55 |
| 1325 | Port Townsend | 48.11167 | -122.757 | 1.19±0.7 | 1.68±0.65 |
| 1347 | Roscoff | 48.7184 | -3.96586 | 1.14±0.45 | 2.48±0.43 |
| 1349 | Lower Escuminac | 47.08333 | -64.8833 | 3.98±0.46 | 3.65±0.67 |
| 135 | Philadelphia (Pier 9n) | 39.93333 | -75.1417 | 4.1±1.05 | 2.28±0.7 |
| 1351 | Raffles Light House | 1.166667 | 103.75 | 3.66±0.51 | 3.18±0.66 |
| 1366 | Tsim Bei Tsui | 22.48722 | 114.0142 | 3.29±1.01 | 3.27±0.65 |
| 1368 | Goods Island | -10.5617 | 142.1517 | 2.5±1.22 | 3.51±1.16 |
| 1386 | Kuchinotsu | 32.60528 | 130.1953 | 2.17±0.61 | 3.21±0.51 |
| 1387 | Maizuru Ii | 35.47667 | 135.3869 | 2.56±0.51 | 3.68±0.66 |
| 1390 | Murotomisaki | 33.26639 | 134.1644 | 7.68±0.83 | 2.42±1.94 |
| 1394 | Point Reyes | 37.995 | -122.977 | 1.66±0.63 | 1.87±0.56 |
| 14 | Helsinki | 60.15363 | 24.95622 | 3.88±1.84 | 4.25±1.73 |
| 1440 | Kaminato Ii (Hatizyo Sima) | 33.13028 | 139.8047 | 4.42±3.99 | 2.96±2.13 |
| 1442 | Hanasaki Ii | 43.27806 | 145.5678 | 0±0.5 | 3.1±0.41 |
| 1446 | Tongyeong | 34.8275 | 128.4347 | 2.33±0.51 | 3.26±0.52 |
| 148 | Baltimore | 39.26667 | -76.5783 | 3.72±0.78 | 2.04±0.8 |
| 1488 | Nagoya Ii | 35.09139 | 136.8808 | 6.42±0.88 | 2.96±0.81 |
| 1489 | Heuksando | 34.68389 | 125.4364 | 1.43±0.57 | 3.44±0.51 |
| 1490 | Ulleung | 37.49139 | 130.9136 | 5.73±1.07 | 5.56±0.78 |
| 1492 | Cape Ferguson | -19.2773 | 147.0584 | 3.94±0.61 | 3.23±0.63 |
| 1493 | Mooloolaba 2 | -26.6833 | 153.1333 | 2.24±0.58 | 3.21±0.46 |
| 1522 | Nase Iii | 28.5 | 129.5 | 2.5±0.73 | 2.97±0.7 |
| 1527 | Gunsan (Outer Port) | 35.97556 | 126.5633 | 4.67±0.73 | 3.37±0.58 |
| 154 | Trieste | 45.64736 | 13.75847 | 1.87±0.62 | 2.4±0.61 |
| 1546 | Geomundo | 34.02833 | 127.3092 | 3.84±0.59 | 3.17±0.52 |
| 1547 | Portland | -38.3434 | 141.6132 | 1.85±0.73 | 1.78±0.53 |
| 1551 | Roompot Buiten | 51.61972 | 3.681944 | 2.01±0.84 | 2.63±0.73 |
| 1568 | Wando | 34.31528 | 126.7597 | 1.97±0.6 | 3.39±0.49 |
| 1585 | Hamada Ii | 34.89722 | 132.0661 | 3.71±0.54 | 2.88±0.55 |
| 1587 | Tappi | 41.2425 | 140.3814 | 2.11±0.47 | 3.16±0.47 |
| 1597 | Rimouski | 48.48333 | -68.5167 | 3.13±0.68 | 3.26±0.84 |
| 161 | Galveston Ii, Pier 21, Tx | 29.31 | -94.7933 | 6.58±1.01 | 4.4±0.8 |
| 1627 | Seogwipo | 33.24 | 126.5617 | 2.55±0.61 | 3.18±0.52 |
| 1629 | Mourilyan Harbour | -17.5833 | 146.0833 | 3.09±0.66 | 3.17±0.75 |
| 1633 | Cherry Point | 48.86333 | -122.757 | 0±0.71 | 1.55±0.68 |
| 1639 | N. Spit, Humboldt Bay | 40.76667 | -124.217 | 5.08±0.67 | 1.54±0.58 |
| 166 | Victoria | 48.41667 | -123.367 | 2.19±0.68 | 1.54±0.65 |
| 167 | Prince Rupert | 54.31667 | -130.333 | 1.18±0.66 | 1.79±0.79 |
| 1671 | Ishigaki Ii | 24.33222 | 124.1636 | 1.72±0.89 | 2.75±0.78 |
| 1673 | Port Louis Ii | -20.15 | 57.5 | 4.36±0.76 | 4.47±0.71 |
| 1674 | Quarry Bay | 22.29111 | 114.2133 | 1.36±0.84 | 3.21±0.65 |
| 1675 | Boryeong | 36.40639 | 126.4861 | 3.59±0.76 | 3.36±0.61 |
| 1699 | Anheung | 36.67361 | 126.1322 | 1.02±0.62 | 3.42±0.59 |
| 1701 | Vaca Key | 24.71167 | -81.105 | 4.34±0.6 | 3.49±0.5 |
| 1704 | Gold Coast Seaway 2 | -27.95 | 153.4167 | 5.51±0.64 | 3.13±0.59 |
| 172 | Mantyluoto | 61.59438 | 21.46343 | 5.15±1.76 | 4.35±1.67 |
| 1745 | Xi Sha | 16.83333 | 112.3333 | 5.38±1.37 | 4.36±0.78 |
| 1748 | Trondheim 2 | 63.43648 | 10.39167 | 3.58±0.72 | 3.49±0.71 |
| 175 | Vancouver | 49.28333 | -123.117 | 1.78±0.71 | 1.48±0.71 |
| 1759 | Viker | 59.03605 | 10.94977 | 3.75±0.94 | 3.85±0.85 |
| 1760 | Rosslyn Bay | -23.161 | 150.7902 | 4±0.54 | 3.2±0.49 |
| 1761 | Hillarys | -31.8256 | 115.7386 | 3.87±0.86 | 2.65±0.86 |
| 1764 | L'estartit | 42.05383 | 3.206448 | 2.02±0.43 | 2.53±0.4 |
| 1789 | Takamatsu Ii | 34.35139 | 134.0569 | 2.33±0.77 | 1.53±0.6 |
| 179 | Smogen | 58.35361 | 11.21778 | 3.42±0.89 | 3.44±0.7 |
| 1790 | Kobe Ii | 34.68222 | 135.1903 | 6.27±0.95 | 0±0.75 |
| 1794 | Workington | 54.65072 | -3.56717 | -1.03±0.86 | 2.72±0.6 |
| 1799 | Winter Harbour | 50.51667 | -128.033 | 1.91±0.78 | 1.38±0.77 |
| 1803 | Tenerife | 28.47722 | -16.2411 | 2.51±0.35 | 2.83±0.3 |
| 1804 | Betio | 1.365056 | 172.933 | 3.47±0.85 | 3.83±0.73 |
| 1805 | Lautoka | -17.6049 | 177.4383 | 2.25±0.61 | 3.55±0.6 |
| 1806 | Bilbao | 43.35153 | -3.04513 | 2.29±0.38 | 1.9±0.4 |
| 1809 | Bonanza | 36.80221 | -6.33813 | 2.41±0.59 | 2.89±0.33 |
| 1810 | Malaga Ii | 36.71184 | -4.41709 | 0.81±0.48 | 2.81±0.42 |
| 1811 | Barcelona | 41.34177 | 2.1657 | 4.62±0.42 | 2.65±0.42 |
| 1813 | Valencia | 39.44203 | -0.31128 | 3.26±0.47 | 2.17±0.4 |
| 183 | Portland (Maine) | 43.65667 | -70.2467 | 3.66±0.58 | 2.94±0.53 |
| 1836 | Lorne | -38.5472 | 143.9888 | 0±0.77 | 2.27±0.57 |
| 1838 | Majuro-C | 7.106028 | 171.3728 | 4.36±0.87 | 4.22±0.83 |
| 1839 | Funafuti B | -8.50247 | 179.1952 | 3.47±0.97 | 4.4±0.88 |
| 1844 | Nauru-B | -0.52942 | 166.9101 | 5.03±1.02 | 4.16±0.72 |
| 1858 | Virginia Key, Fl | 25.73 | -80.1617 | 4.82±0.72 | 3.51±0.52 |
| 188 | Key West | 24.555 | -81.8067 | 3.93±0.62 | 3.47±0.53 |
| 189 | Port Hedland | -20.3176 | 118.5744 | 2.06±0.91 | 3.06±0.85 |
| 1898 | Vigo Ii | 42.24314 | -8.726 | 1.48±0.57 | 2.73±0.39 |
| 194 | Pietarsaari / Jakobstad | 63.70857 | 22.68958 | 4.9±1.82 | 4.5±1.71 |
| 20 | Vlissingen | 51.44222 | 3.596111 | 2.48±0.85 | 2.61±0.72 |
| 203 | Furuogrund | 64.91583 | 21.23056 | 5.02±1.85 | 4.62±1.72 |
| 2072 | Port Alma | -23.5833 | 150.8667 | 2.94±0.61 | 3.32±0.52 |
| 2074 | Bowen Ii | -20.0167 | 148.25 | 3.19±0.57 | 3.41±0.54 |
| 2101 | Kalix | 65.69694 | 23.09611 | 5.05±1.94 | 4.94±1.83 |
| 2103 | Forsmark | 60.40861 | 18.21083 | 4.69±1.65 | 4.17±1.63 |
| 2105 | Visby | 57.63917 | 18.28444 | 4.33±1.53 | 3.77±1.61 |
| 2106 | Oskarshamn | 57.275 | 16.47806 | 4.28±1.51 | 3.67±1.37 |
| 2107 | Simrishamn | 55.5575 | 14.35778 | 3.29±1.34 | 3.53±1.19 |
| 2108 | Skanor | 55.41667 | 12.82944 | 3.72±1.23 | 3.36±1.01 |
| 2109 | Barseback | 55.75639 | 12.90333 | 3.88±1.09 | 3.32±0.98 |
| 2110 | Viken | 56.14222 | 12.57917 | 3.57±1.05 | 3.34±0.91 |
| 2111 | Ringhals | 57.24972 | 12.1125 | 3.21±0.99 | 3.72±0.8 |
| 2112 | Stenungsund | 58.09333 | 11.8325 | 3.86±1.04 | 3.63±0.82 |
| 2113 | Kungsvik | 58.99667 | 11.12722 | 4.2±0.97 | 3.82±0.81 |
| 2127 | Port Angeles, Washington | 48.125 | -123.44 | 1.39±0.77 | 1.58±0.68 |
| 216 | Port Pirie | -33.1776 | 138.0117 | 4.5±0.86 | 2.4±0.86 |
| 22 | Hoek Van Holland | 51.9775 | 4.12 | 1.42±0.92 | 2.69±0.79 |
| 224 | Lewes (Breakwater Harbor) | 38.78167 | -75.12 | 3.71±0.77 | 2.21±0.63 |
| 225 | Ketchikan | 55.33167 | -131.625 | 0±0.66 | 1.49±0.64 |
| 229 | Kemi | 65.67337 | 24.51525 | 4.45±1.93 | 4.99±1.9 |
| 2295 | Beaufort, North Carolina | 34.72 | -76.67 | 3.65±0.73 | 4.55±0.79 |
| 23 | Den Helder | 52.96444 | 4.745 | 1.59±1.07 | 2.8±0.88 |
| 230 | Port Lincoln | -34.7159 | 135.87 | 0±0.83 | 2.53±0.69 |
| 2310 | Brunswick Heads | -28.5368 | 153.5521 | 1.8±0.71 | 3.14±0.58 |
| 2324 | Lewisetta, Virginia | 37.995 | -76.465 | 5.37±0.8 | 1.94±0.7 |
| 2325 | Oregon Inlet Marina, North Carolina | 35.795 | -75.5483 | 4.38±0.76 | 4.37±0.82 |
| 2330 | Port Chicago, California | 38.055 | -122.04 | 1.63±0.98 | 1.9±0.59 |
| 234 | Charleston I | 32.78167 | -79.925 | 4.61±0.92 | 3.94±0.81 |
| 236 | West-Terschelling | 53.36306 | 5.22 | 3.38±1.16 | 2.8±0.91 |
| 239 | Turku / Abo | 60.42828 | 22.10053 | 4.07±1.76 | 4.23±1.68 |
| 24 | Delfzijl | 53.32639 | 6.933056 | 0±1.45 | 2.89±0.93 |
| 240 | Raahe / Brahestad | 64.66633 | 24.40705 | 4.38±1.87 | 4.75±1.76 |
| 249 | Foglo / Degerby | 60.03188 | 20.38482 | 3.98±1.69 | 4.14±1.65 |
| 25 | Harlingen | 53.17556 | 5.409444 | 3.06±1.41 | 2.81±0.92 |
| 256 | La Jolla (Scripps Pier) | 32.86667 | -117.257 | 0.73±0.58 | 1.68±0.61 |
| 265 | Astoria (Tongue Point) | 46.20667 | -123.768 | 0±0.96 | 1.61±0.86 |
| 285 | Kaskinen / Kasko | 62.34395 | 21.21483 | 4.65±1.76 | 4.37±1.7 |
| 299 | Sewells Point, Hampton Roads | 36.94667 | -76.33 | 4.84±0.87 | 3.34±0.54 |
| 302 | Tregde | 58.00638 | 7.554759 | 3.09±0.49 | 3.36±0.79 |
| 310 | Yamba | -29.4297 | 153.3619 | 2.47±0.83 | 3.07±0.57 |
| 313 | Heimsjoen | 63.42522 | 9.101504 | 5±0.72 | 3.38±0.6 |
| 315 | Hamina | 60.56277 | 27.1792 | 3.92±2.02 | 4.22±1.85 |
| 32 | Ijmuiden | 52.46222 | 4.554722 | 0±1.03 | 2.77±0.86 |
| 33 | Oscarsborg | 59.67807 | 10.60486 | 3.81±1.05 | 3.44±0.71 |
| 330 | Klagshamn | 55.52222 | 12.89361 | 1.83±1.19 | 3.35±1 |
| 332 | Eastport | 44.90333 | -66.9817 | 3.8±0.45 | 2.86±0.5 |
| 351 | Newport | 41.505 | -71.3267 | 3.53±0.53 | 3.05±0.52 |
| 353 | Bakar | 45.305 | 14.54 | 0±0.67 | 2.44±0.63 |
| 366 | Sandy Hook | 40.46667 | -74.0083 | 4.63±0.73 | 2.65±0.65 |
| 367 | Woods Hole (Ocean. Inst.) | 41.52333 | -70.6717 | 4.17±0.53 | 3.03±0.49 |
| 376 | Rauma / Raumo | 61.13353 | 21.42582 | 4.08±1.73 | 4.29±1.66 |
| 377 | Santa Monica (Municipal Pier) | 34.00833 | -118.5 | 2.38±0.57 | 1.84±0.56 |
| 384 | Friday Harbor (Ocean. Labs.) | 48.54667 | -123.01 | 1.91±0.68 | 1.63±0.63 |
| 385 | Neah Bay | 48.36667 | -124.612 | 0±0.79 | 1.34±0.84 |
| 393 | St. John's, Nfld. | 47.56667 | -52.7167 | 1.61±0.75 | 3.42±0.45 |
| 395 | Fort Pulaski | 32.03333 | -80.9017 | 5.47±0.95 | 3.84±0.88 |
| 396 | Wilmington | 34.22667 | -77.9533 | 4.85±1.05 | 4.25±0.81 |
| 397 | Sassnitz | 54.51083 | 13.64306 | 3.2±1.3 | 3.39±1.18 |
| 413 | Oostende | 51.23333 | 2.916667 | 2.02±0.78 | 2.59±0.69 |
| 427 | Charlottetown | 46.23333 | -63.1167 | 3.43±0.58 | 3.36±0.38 |
| 429 | New London | 41.36 | -72.09 | 3.98±0.62 | 2.92±0.54 |
| 437 | Alameda (Naval Air Station) | 37.77167 | -122.298 | 0.77±0.73 | 1.93±0.51 |
| 448 | Port Adelaide (Outer Harbor) | -34.7798 | 138.4807 | 2.49±0.88 | 2.39±0.63 |
| 453 | Le Havre | 49.4819 | 0.106 | 2.11±0.52 | 2.46±0.49 |
| 467 | Cherbourg | 49.6513 | -1.63563 | 1.4±0.43 | 2.36±0.43 |
| 47 | Stavanger | 58.97434 | 5.730121 | 2.91±0.56 | 3.33±0.48 |
| 486 | Maloy | 61.93378 | 5.11331 | 4.37±0.68 | 3.28±0.54 |
| 489 | Nieuwpoort | 51.15 | 2.733333 | 0.91±0.76 | 2.58±0.7 |
| 497 | Port Isabel | 26.06 | -97.215 | 4.87±0.86 | 4.65±0.69 |
| 508 | Port San Luis | 35.17667 | -120.76 | 2.82±0.56 | 2.15±0.55 |
| 509 | Alesund | 62.46941 | 6.151946 | 3.54±0.73 | 3.17±0.6 |
| 518 | Kushiro | 42.97556 | 144.3714 | 4.5±0.39 | 2.95±0.45 |
| 520 | St. Petersburg | 27.76 | -82.6267 | 5.62±0.52 | 4.64±0.53 |
| 526 | Grand Isle | 29.26333 | -89.9567 | 3.33±0.74 | 4.82±0.78 |
| 562 | Bodo | 67.28829 | 14.39081 | 6.05±0.71 | 3.98±0.75 |
| 563 | Gibara | 21.10833 | -76.125 | 2.79±0.7 | 3.24±0.52 |
| 564 | Mackay | -21.1 | 149.2333 | 3.25±0.63 | 3.44±0.57 |
| 571 | Talcahuano | -36.6833 | -73.1 | -9.72±0.71 | 2.14±0.51 |
| 58 | Bergen | 60.39805 | 5.320487 | 3.27±0.56 | 3.4±0.52 |
| 62 | Oslo | 59.90856 | 10.73451 | 3.84±1.08 | 3.46±0.74 |
| 636 | Kiptopeke Beach | 37.165 | -75.9883 | 3.09±0.76 | 3.21±0.55 |
| 637 | Townsville I | -19.25 | 146.8333 | 3.68±0.62 | 3.2±0.69 |
| 682 | Kristiansund | 63.11386 | 7.734352 | 3.27±0.72 | 3.37±0.6 |
| 683 | Burnie | -41.0501 | 145.915 | 1.31±0.38 | 2.28±0.51 |
| 69 | Olands Norra Udde | 57.36611 | 17.09722 | 3.74±1.56 | 3.75±1.5 |
| 7 | Cuxhaven 2 | 53.86667 | 8.716667 | 2.15±1.81 | 2.94±0.95 |
| 70 | Kungsholmsfort | 56.10528 | 15.58944 | 4.03±1.44 | 3.56±1.29 |
| 701 | Kainan | 34.14417 | 135.1914 | 2.48±0.82 | 3.05±0.63 |
| 722 | Yokosuka | 35.28806 | 139.6514 | 10.22±0.6 | 5.12±0.83 |
| 723 | Dalian | 38.86667 | 121.6833 | 4.79±0.68 | 3.34±0.6 |
| 724 | Sembawang | 1.466667 | 103.8333 | 3.25±0.51 | 3.22±0.5 |
| 754 | Lowestoft | 52.47311 | 1.750111 | 2.75±0.83 | 2.71±0.71 |
| 78 | Stockholm | 59.32417 | 18.08167 | 4.32±1.6 | 4.05±1.63 |
| 79 | Oulu / Uleaborg | 65.04032 | 25.41823 | 4.81±1.91 | 4.95±1.84 |
| 809 | Komatsushima | 34.00917 | 134.5878 | 2.74±0.83 | 2.44±0.62 |
| 811 | Sakai | 35.54778 | 133.2431 | 2.51±0.55 | 2.8±0.56 |
| 812 | Uno | 34.48861 | 133.9494 | 4.26±0.79 | 2.04±0.56 |
| 813 | Hakodate I | 41.78167 | 140.7247 | 2.61±0.42 | 3.24±0.41 |
| 814 | Aburatsu | 31.57694 | 131.4094 | 2.03±0.69 | 3.07±0.62 |
| 815 | Tosa Shimizu | 32.77917 | 132.9589 | 2.89±0.71 | 2.13±1.13 |
| 816 | Wakayama | 34.22167 | 135.1456 | 3.26±0.84 | 1.98±0.65 |
| 817 | Tan-Nowa | 34.33889 | 135.1781 | 3.56±0.81 | 1.03±0.69 |
| 819 | Mar Del Plata (Club) | -38.0333 | -57.5167 | 3.93±0.87 | 3.05±0.43 |
| 822 | Brisbane (West Inner Bar) | -27.3667 | 153.1667 | 3.51±0.6 | 2.81±0.59 |
| 824 | Petropavlovsk-Kamchatsky | 52.98333 | 158.65 | 6.59±0.63 | 2.14±0.51 |
| 829 | Queen Charlotte City | 53.25 | -132.067 | 3.09±0.56 | 1.31±0.65 |
| 832 | Palermo | -34.5667 | -58.4 | 4.9±1.24 | 2.81±1.02 |
| 833 | Eden | -37.0737 | 149.9077 | 0.72±0.56 | 3.06±0.47 |
| 834 | Bunbury | -33.3234 | 115.66 | 1.69±0.85 | 2.81±0.93 |
| 835 | Karumba | -17.5 | 140.8333 | 2.54±1.36 | 3.74±1.49 |
| 88 | Ratan | 63.98611 | 20.895 | 5.06±1.79 | 4.49±1.7 |
| 9 | Maassluis | 51.9175 | 4.249722 | 1.68±0.94 | 2.72±0.83 |
| 933 | Zhapo | 21.58333 | 111.8167 | 2.74±0.85 | 3.53±0.66 |
| 934 | Kanmen | 28.08333 | 121.2833 | 4.85±0.91 | 3.11±0.69 |
| 935 | Darwin | -12.4718 | 130.8459 | 3.5±0.87 | 3.68±0.89 |
| 953 | Cairns | -16.9167 | 145.7833 | 0.87±0.62 | 3.11±0.71 |
| 954 | Mokpo | 34.77972 | 126.3756 | 6.33±0.77 | 3.39±0.51 |
| 955 | Busan | 35.09611 | 129.0356 | 2.62±0.5 | 3.29±0.58 |
| 982 | Devonport | 50.36839 | -4.18525 | 1.23±0.52 | 2.44±0.39 |
| 984 | Bella Bella | 52.16667 | -128.133 | 2.25±0.72 | 1.26±0.81 |
| 996 | Port Taranaki | -39.0562 | 174.0344 | 1.91±0.73 | 4.2±0.46 |